\documentclass[aps,prb,twocolumn,showpacs,preprintnumbers,amsmath,amssymb]{revtex4-2}

\usepackage[version=3]{mhchem} 
\usepackage[T1]{fontenc}       
\usepackage{miller}
\usepackage[utf8]{inputenc}
\usepackage{graphicx}
\usepackage{amsfonts}
\usepackage{epstopdf}
\usepackage{float}

\usepackage{gensymb}
\usepackage{datetime}
\usepackage{color}

\usepackage[english]{babel}

\usepackage[separate-uncertainty=true,detect-all,retain-explicit-plus]{siunitx}

\setcitestyle{super}

\begin{document}

\title{Template and Temperature Controlled Polymorph Formation in Squaraine Thin Films}

\author{Frank Balzer}
\affiliation{SDU Centre for Photonics Engineering, University of Southern Denmark, DK-6400 S{\o}nderborg, Denmark}

\author{Tobias Breuer}
\affiliation{Department of Physics, Philipps University of Marburg, D-35032 Marburg, Germany}

\author{Gregor Witte}
\affiliation{Department of Physics, Philipps University of Marburg, D-35032 Marburg, Germany}

\author{Manuela Schiek}
\email{manuela.schiek@jku.at}
\affiliation{Institute of Physics, University of Oldenburg, D-26111 Oldenburg, Germany}
\affiliation{Center for Surface- and Nanoanalytics (ZONA) \& Linz Institute for Organic Solar Cells (LIOS), Johannes Kepler University, A-4040 Linz, Austria}


\begin{abstract}
Controlling the polymorph formation in organic semiconductor thin films by the choice of substrate and deposition temperature is a key factor for targeted device performance. Small molecular semiconductors such as the quadrupolar donor-acceptor-donor (D-A-D) type squaraine compounds allow both solution and vapor phase deposition methods. A prototypical anilino squaraine with branched butyl chains as terminal functionalization (SQIB) has been considered for photovoltaic applications due to its broad absorption within the visible to deep-red spectral range. Its opto-electronic properties depend on the formation of the two known polymorphs adopting a monoclinic and orthorhombic crystal phase. Both phases emerge with a strongly preferred out-of-plane and rather random in-plane orientation in spincasted thin films depending on subsequent thermal annealing. Upon vapor deposition on dielectric and conductive substrates, such as silicon dioxide, potassium chloride, graphene and gold, the polymorph expression depends on the choice of growth substrate. In all cases the same pronounced out-of-plane orientation is adopted, but with a surface templated in-plane alignment in case of crystalline substrates. Combining X-ray diffraction, atomic force microscopy, ellipsometry and polarized spectro-microscopy we identify the processing dependent evolution of the crystal phases, correlating morphology and molecular orientations within the textured SQIB films.

\smallskip \noindent \textbf{Version:} \today, \currenttime
\end{abstract}

\maketitle

\section{Introduction}
\begin{figure}[th]
\centering
\includegraphics[width=0.47\textwidth]{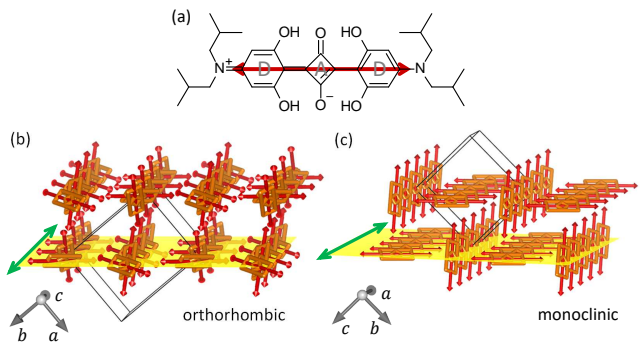}
\caption{(a) Structural formula of the donor-acceptor-donor type (D-A-D) SQIB molecule. The red arrow denotes the long molecular axis and the direction of the transition dipole moment for the $S_0 \rightarrow S_1$ transition. Packing and stacking of SQIB molecules and the observed planes parallel to the surface for the two known polymorphs are depicted in (b) and (c) using VESTA.\cite{Momma11} The primitive unit cells are denoted by thin black lines. The long molecular axes of the SQIB molecules, and with that the transition dipole moments, are shown by red arrows. Molecular backbones are depicted by orange rectangles to visualize their stacking. The molecular stacking directions are indicated by green arrows pointing parallel to the crystallographic $c$-axis and $a$-axis in (b) and (c), respectively.
The yellow planes denote the \hkl(110) plane of the orthorhombic $Pbcn$ phase ($Z$ = 4) (b), and the \hkl(011) plane of the monoclinic $P21/c$ phase ($Z$ = 2) (c), respectively. }
\label{sqib_packing}
\end{figure}

Crystalline organic thin films often exhibit linear dichroism and birefringence.\cite{Kahr2010} By structural design of the molecular building blocks, advanced functionality can be introduced including nonlinear optical properties\cite{Brewer08} or circular dichroism.\cite{Zablocki20,Schulz18} In addition, fine-tuning can be obtained by selecting a specific polymorph through the processing conditions.\cite{Llinas08,Gentili19,Jones16,CruzCabeza15}
The interest in selective polymorph growth of organic molecules is widespread, since the polymorph choice can improve opto-electronic device performance,\cite{Lee11,Bischof2021} or decide about the bio-functionality of for instance active drug ingredients.\cite{Blagden2007,Hilfiker18,Lee14,Yang2017}
The dihydroxy anilino squaraine SQIB (2,4-bis[4-(\textit{N,N}-diisobutylamino)-2,6-dihydroxy\-phenyl]squar\-aine), as sketched in Figure~\ref{sqib_packing}(a), is a prototypical quadrupolar donor-acceptor-donor type semiconductor compound. Such squaraines have been widely used for various photovoltaic application scenarios due to their intense interaction with visible to deep-red light. Applications include xerography,\cite{Halton08,Weiss16} solar cells,\cite{Chen15b,Chen19,Brueck14,Scheunemann17,Scheunemann2019,Chen2018} photodetectors\cite{Schulz19} and neurostimulating photo-capacitors.\cite{Abdullaeva18,Balzer19}

In case of SQIB two polymorphs are known, each having multiple molecules in the primitive unit cell:\cite{Balzer17a,Viterisi2014} a monoclinic $P21/c$ phase ($Z$ = 2; CCDC code 1567209), and an orthorhombic $Pbcn$ phase ($Z$ = 4; CCDC code 1567104), as depicted in Figures~\ref{sqib_packing}(b) and (c).
The optical properties of the condensed phases are dominated by Coulombic coupling between the molecular compounds. In a simplified picture according to the Kasha model the excitonic properties can be described by linear combinations of multiple transition dipole moments.\cite{Davydov64,Kasha65,Hestand18,Breuer2012,Funke21} Here, the transition dipole moment is along the molecular backbone as depicted by red arrows in Figure~\ref{sqib_packing}. For molecular solids with nonprimitive basis, the coupling of the different molecules causes a distinct excitonic spectral  splitting of the absorption band (Davydov splitting) into an upper and a lower Davydov component, UDC and LDC, respectively. The pronounced coupling between equivalent molecules forming stacks cause an overall spectral shift of both Davydov components relative to the monomer absorbance in solution, as discussed previously in reference \citenum{Balzer17a} and for completeness shown in the Supporting Information in Figure~S1. With that, the monoclinic polymorph can be described as an H-type aggregate (overall spectral blueshift) and the orthorhombic polymorph as a J-type aggregate (overall spectral red-shift).

The polarized absorbance properties in normal incidence transmission of crystalline textured thin film samples are then determined by the projection of the directions of the Davydov transitions onto the substrate plane.\cite{Balzer17a} The yellow planes in Figures~\ref{sqib_packing}(b) and (c) visualize the molecular arrangement within the \hkl(110) plane and the \hkl(011) plane, which have been observed previously for the two SQIB polymorphs spincasted on non-templating glass to be parallel to the surface. Here, the polymorph formation is controlled by thermal post-annealing of the samples as discussed earlier.\cite{Balzer17a,Freese18}

In other work, also the solvent and amorphous coatings on the substrates have been found to be influential on the polymorph formation in addition to annealing temperature but without changing the out-of-plane orientation.\cite{Viterisi2014} Those results have been obtained for spincasting of blended solutions containing a soluble fullerene acceptor such as PCBM, which has not altered crystallization propensity and crystallographic orientation. This is consistent with our results for spincasting PCBM-blended solutions on substrates with amorphous coatings such as \ce{MoO3}\cite{Scheunemann17,Scheunemann2019} and PEDOT:PSS\cite{Scheunemann2019} typically used as interfacial layers in photovoltaic devices. Subsequent thermal annealing at \SI{180}{\celsius} of a SQIB:PCBM blend with 1:1 ratio by weight spincasted on Indium Tin Oxide (ITO) results in concomitant polymorphs but favoring the orthorhombic one.\cite{Abdullaeva18,Balzer19} Detailed inspections by AFM and TEM cross-section imaging revealed the formation of a bilayer structure.\cite{Balzer19} A phase separation happens during annealing, and the PCBM sinks to the bottom leaving characteristic elongated holes behind in the orthorhombic platelets forming on top. See also Table~\ref{summary} for an overview of previously obtained results.

SQIB is one among the rare examples of donor-type semiconductors that can equifeasible be deposited from solution and by thermal vapor deposition. Also co-deposition with a fullerene acceptor such as C60 and C70 is possible, thereby, allowing to study photovoltaic device performance comparing solution and vacuum-processed bulk-heterojunctions.\cite{Chen2018} For both processing strategies, the amorphous SQIB phase was favored using low thermal post-annealing temperatures and substrate temperatures during vapor deposition of the non-templating \ce{MoO3} und PEDOT:PSS interfacial layers.

In this work, the focus is on polymorph selection, (out-of-plane) orientation and (in-plane) alignment control via templating of vapor deposited SQIB. For this purpose, SQIB films have been deposited via organic molecular beam deposition (OMBD) on various dielectric and conductive substrates at elevated temperatures: silicon dioxide, potassium chloride, graphene and gold. Thereby we show, that vapor phase heterogeneous nucleation on different substrates (templating) allows polymorph selection.
By contrast, solution-processed SQIB films are initially amorphous but allow a temperature-controlled polymorph selection by a post-deposition thermal annealing procedure. As intramolecular interactions exceed the molecule-substrate interactions the same film orientations occur for the two polymorphs formed in the annealed and vapor deposited films, while the in-plane alignment is to some extend directed by the substrate.

\section{Results and Discussion}
SQIB thin films with nominal thicknesses of \SI{30}{nm} have been obtained by OMBD on \ce{SiO2} (silicon wafer with native oxide), on a \hkl(111) gold layer supported on a mica substrate, on a graphene layer supported on a quartz substrate, and on a cleaved \hkl(001) face of a \ce{KCl} single crystal. To enhance the crystalline order, the substrates have been heated to \SI{100}{\celsius} during deposition, except for \ce{KCl} which was heated to \SI{120}{\celsius}. Note that whereas some squaraine molecules decompose during evaporation, SQIB is thermally stable because of the intermolecular hydrogen bonds from the hydroxy groups at the anilino ring and the squaric oxygen.\cite{Tian02,Chen2018} We have verified this by near edge X-ray absorption fine structure spectroscopy (NEXAFS), see Figure~S2 in the Supporting Information, showing that the characteristic NEXAFS signature is identical for evaporated films and the raw powder.
For comparison, additional SQIB films were prepared by spincasting of chloroform solutions onto glass substrates, and the temperature induced crystallization was monitored by polarized optical microscopy.

\begin{figure}[t]
\centering
\includegraphics[width=0.49\textwidth]{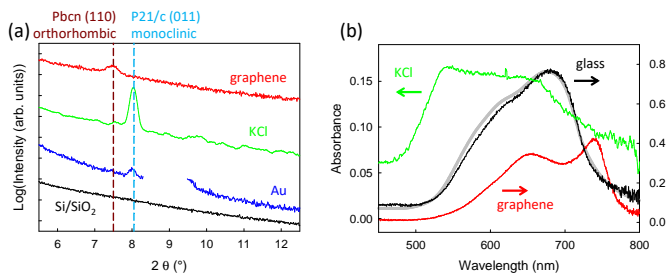}
\caption{(a) Specular X-ray diffractograms are measured with \ce{Cu} K$\alpha$ radiation of SQIB on various substrates. Both polymorphs are found: the orthorhombic $Pbcn$ and the monoclinic $P21/c$ phase depending on the growth substrate. The calculated positions for the \hkl(110) and \hkl(011) reflections are shown as dashed vertical lines. Note that for SQIB on \ce{Au}\hkl(111)/mica, a dominating mica peak at $2\theta \approx \SI{9}{\degree}$ was omitted. Vapor-deposited SQIB on \ce{SiO2} as well as non-annealed, spincasted films on glass (not shown) are XRD-silent.
(b) Absorbance spectra measured in normal incidence transmission of non-annealed spincasted (glass) and evaporated (graphene, \ce{KCl}) SQIB thin films. The gray line is the absorbance calculated based the complex refractive index obtained from amorphous SQIB on \ce{SiO2} by ellipsometry as shown in Figure~\ref{SQIB_nk_iso}(b). }
\label{xrd}
\end{figure}

Specular X-ray diffractograms, Figure~\ref{xrd}(a), allow to identify the adopted polymorph and the orientation of the films on various substrates. Vapor-deposited SQIB layers on \ce{SiO2} (black line) as well as non-annealed, spincasted SQIB films on glass (not shown in the graph, see \citenum{Balzer17a} for reference) are XRD-silent indicating an amorphous isotropic phase. For the crystalline phases two diffraction peaks can be identified, one at $2\theta=\SI{7.6}{\degree}$ corresponding to the $Pbcn$ \hkl(110) plane, and another one at $2\theta=\SI{8.1}{\degree}$ corresponding to the $P21/c$ \hkl(011) plane.\cite{Balzer17a} On \ce{Au}\hkl(111)/muscovite mica (blue line), \hkl(011) oriented films of the monoclinic phase are formed, while on graphene (red line) the orthorhombic phase with \hkl(110) orientation is clearly prevailing. On \ce{KCl}\hkl(001) both polymorphs can be identified but the monoclinic phase with \hkl(011) orientation is dominating (green line).
For both realized crystallographic orientations, the molecular stacking direction (\hkl[001] for the orthorhombic polymorph, \hkl[100] for the monoclinic one) is parallel to the substrate, see also Figure~\ref{sqib_packing}. This is caused by the strong intermolecular interactions given by the slipped-$\pi$-stacking of the D-A-D backbones. These interactions in the crystalline bulk thin films exceed the molecule-substrate interactions on the presently investigated inert substrates. Note that for a submonolayer coverage of SQIB molecules on a silver surface a wetting layer dictated by molecule-substrate interactions has been identified.\cite{Luft18}
Interestingly, we have detected that the polymorph can be selected by the choice of growth substrate, but reveal only distinct orientations on all investigated substrates ensuring the molecular stacking direction to run parallel to the surface.

In Figure~\ref{xrd}(b) the unpolarized absorbance spectra ($Abs=-\log T$, $T$ normal incidence transmission) of the two SQIB polymorphs with their specific orientation as well as of an amorphous, isotropic SQIB film are summarized. The amorphous phase on glass (black line) has a broad absorbance peaking at \SI{670}{nm} with a vibronic shoulder at \SI{615}{nm}. The latter is in close agreement with the spectrum calculated (grey line) using the complex refractive index obtained by ellipsometry from an evaporated film on \ce{SiO2}. The \hkl(011)-orientated monoclinic phase found on \ce{KCl} has a blue-shifted absorbance with a broad absorption from \SI{530}{nm} to \SI{630}{nm} (green line). The absorbance is red-shifted for the \hkl(110)-oriented orthorhombic polymorph found on graphene with a well resolved Davydov splitting of \SI{0.23}{eV}, peaking at \SI{652}{nm} and \SI{740}{nm} (red line). Due to the pronounced and characteristic differences of the spectra, simple UV/Vis spectroscopy allows to distinguish conclusively between the amorphous and the two crystalline SQIB phases. Polarized spectro-microscopy then allows mapping of samples containing concomitant polymorphs, and adds information about the local in-plane orientation. In the following, the impact of the substrate on the polymorph formation and the alignment is elucidated in more detail. Initially, a fresh \textit{in-situ} perspective on already known post-annealing polymorph control of solution-processed films is presented.

\subsection{Spincoated SQIB on Glass}

\begin{figure}[t]
\centering
\includegraphics[width=0.48\textwidth]{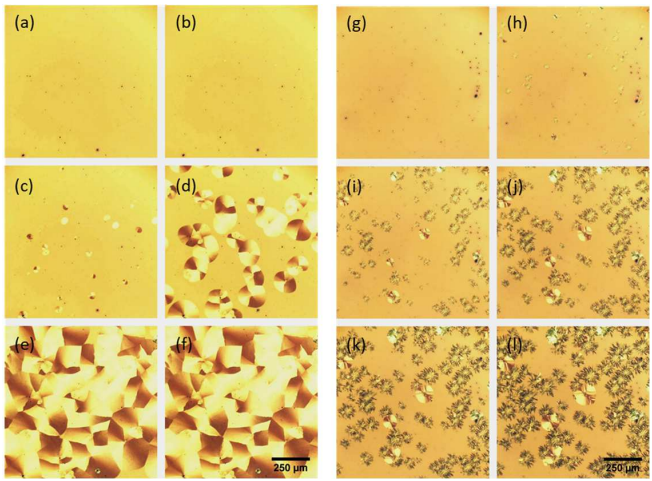}
\caption{Polarized optical reflection microscope images (single polarizer) of orthorhombic phase platelets (a-f) and monoclinic phase crystallites (g-l) formation during heating. For the platelets, images have been extracted from the movie at times (a) \SI{2.4}{s}, (b) \SI{2.9}{s}, (c) \SI{3.5}{s}, (d) \SI{4.1}{s}, (e) \SI{4.6}{s}, and (f) \SI{5.2}{s} after placing the sample on a hotplate with a surface temperature of \SI{195}{\celsius}. For the monoclinic phase crystallite formation, similar snapshots are taken  at the times (g) \SI{4}{min}, (h) \SI{15}{min}, (i) \SI{26}{min}, (j) \SI{37}{min}, (k) \SI{49}{min}, and (l) \SI{60}{min} after placing the sample on a hotplate with a surface temperature of \SI{80}{\celsius}. }
\label{moviestills}
\end{figure}

Spincasted thin films of neat SQIB on bare and indium tin oxide (ITO) coated glass have already been documented before.\cite{Balzer17a,Balzer19,Funke21} Chloroform was used as a rapidly evaporating solvent. Upon subsequent thermal annealing, depending on the annealing temperature, the two previously described polymorphs are found to various extents. Such samples have already been well characterized by XRD and polarized spectro-microscopy after finishing the annealing step. Here, we monitor \textit{in situ} the polymorph formation process during annealing at two characteristic temperatures. Movies from both crystallization processes can be found in the Supporting Information section.
In Figures~\ref{moviestills}(a-f), a time series of polarized optical micrographs extracted from the movies are shown. At the beginning, the amorphous SQIB film spincasted on glass is placed on a hotplate preheated to a surface temperature of \SI{195}{\celsius}. Within seconds, platelets of the orthorhombic polymorph appear. When those start touching each other, domain boundaries are formed. AFM scans across a domain boundary between adjacent domains reveal a characteristic gap and allow to determine the effective film/platelet thickness of around \SI{50}{nm} for this sample, see Figure~S3 in the Supporting Information section. After less than ten seconds, the amorphous film transforms completely into a\hkl(110)-oriented polycrystalline film.

The extended orthorhombic phase domains are not single crystalline, but yet, they are well suited for polarized spectro-microscopy investigation as previously published.\cite{Balzer17a,Balzer19} For completeness, such an analysis is shown in the Supporting Information in Figures~S4 and S5 for transmission and reflection. The complete diagonal dielectric tensor of SQIB could be determined by imaging ellipsometry, well reproducing the Davydov splitting.\cite{Funke21} For completeness, the complex refractive indices along the three crystallographic axes are shown in the Supporting Information in Figure~S6.

Placing a similar sample onto a hotplate with a lower temperature such as \SI{80}{\celsius}, Figures~\ref{moviestills}(g-l), fibrous crystallite domains of the monoclinic polymorph evolve growing from randomly distributed seeds. Note that also here a few orthorhombic phase platelets form. However, the timescale is much larger than for formation of the orthorhombic polymorph in Figures~\ref{moviestills}(a-f). The whole process lasts several hours until the entire, initially amorphous film is crystallized.

For intermediate annealing temperatures, both polymorphs crystallize to various extents on the surface. That way, spincasted thin films can be crystallized into the desired polymorph or into specific ratios of concomitant polymorphs steered by the post-annealing temperature.

This selective recrystallization is at least true for using chloroform as solvent for casting of the SQIB films. This rapidly evaporating solvent gives little time for the molecules to arrange into an equilibrium state.\cite{Cranston2021} In these amorphous films, voids appear due to evaporated solvents, which allow for reorganization and thermally activated recrystallization.
Furthermore, the crystalline domains size is regulated by (random) occurrence of crystallization seeds, such colloidal aggregates within the solution of dust particles. However, the preferred crystalline orientation always remains the same for each polymorph driven by intermolecular interactions.

\subsection{OMBD of SQIB on \ce{SiO2}}

\begin{figure}[t]
\centering
\includegraphics[width=0.48\textwidth]{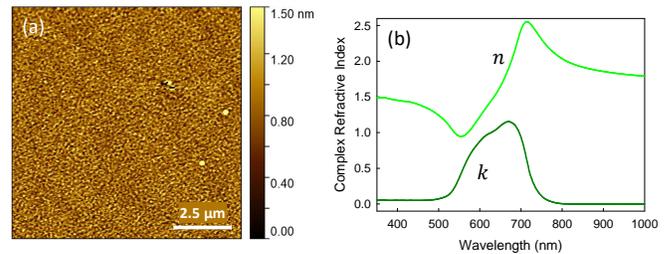}
\caption{(a) AFM image and (b) real and imaginary part of the complex index of refraction, $N=n + i k$, determined by variable angle spectroscopic ellipsometry of a SQIB film vapor deposited on \ce{SiO2} (\ce{Si}-wafer covered with native oxide).}
\label{SQIB_nk_iso}
\end{figure}

For the vacuum deposition of SQIB on a \ce{Si}-wafer with \SI{2}{nm} native oxide at \SI{100}{\celsius} the XRD is featureless indicating that an amorphous thin film is formed, Figure~\ref{xrd}(b). Atomic force microscopy, Figure~\ref{SQIB_nk_iso}(a), reveals that the film with a nominal thickness of \SI{30}{nm} as derived from the Quartz microbalance is very smooth with a RMS roughness of \SI{0.3}{nm}. Therefore, X-ray reflectivity (XRR) scans show well resolved Kiessig fringes at small scattering angles, see Supporting Information Figure~S7, which confirms the low surface roughness inferred by AFM. From such a smooth and extended thin SQIB film the complex refractive index $N=n+i k$ could easily be determined by standard spectroscopic ellipsometry confirming its isotropic nature, Figure~\ref{SQIB_nk_iso}(b). Two samples with different layer thickness have been fitted simultaneously for improved reliability of the fit.\cite{Zablocki2020c} For the sample shown in Figure~\ref{SQIB_nk_iso}(a) ellipsometry determines the thickness to be \SI{25}{nm}.
A comparison between measured and calculated reflection spectra using OpenFilters\cite{Larouche08} for various angles of incidence shows a good agreement, see Supporting Information Figure~S8. Likewise, the absorbance of non-annealed spincoated SQIB films on glass is reproduced satisfactorily, Figure~\ref{xrd}(b). Interestingly, a vapor deposited SQIB thin film on a silicon dioxide substrate has been reported by others to have a slight out-of-plane anisotropy.\cite{Chen12b} This hints to a dependence on processing parameters such as substrate temperature and deposition rate on the film formation.
However, in both cases the maximum of the extinction coefficient $k$ is large, easily exceeding one. With that, such amorphous SQIB thin films are among the top candidates for light-harvesting organic photovoltaic materials.\cite{Vezie16} Even though all extinction coefficient values of the anisotropic tensor of the crystalline orthorhombic phase SQIB are larger by a factor of the roughly three,\cite{Funke21} see above and Supporting Information Figure~S6, the amorphous, isotropic phase is favored for macroscopic light-harvesting device application. This is due to the absence of polarization dependent absorbance as well as domain boundaries as barriers for charge carriers.\cite{Zheng2020} Nevertheless, local anisotropy can be beneficial for microscopic applications such as photovoltaic stimulation of living cells.\cite{Abdullaeva18,Balzer19}

Most remarkably, the vapor deposited SQIB films on silicon are very stable in their amorphous isotropic phase in the present study. Neither growth at medium-elevated substrate temperatures, nor post-annealing after deposition of the sample does induce a notable crystallization. This is distinctly different for spincoated films prepared at ambient temperatures on glass as discussed above. There, rapid evaporation of the solvent used for spincasting is suspected to leave voids for subsequent thermally activated molecular reorganization. Here, in case of vapor deposition the molecules have more time to arrange into a densely packed solid films leaving no room for reorganization.\cite{Cranston2021}

\subsection{OMBD of SQIB on Graphene/Quartz}

\begin{figure}[h]
\centering
\includegraphics[width=0.48\textwidth]{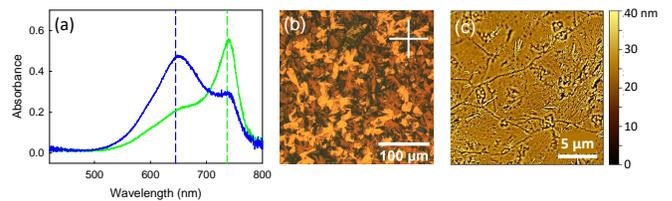}
\caption{(a) Polarized absorbance spectra of a single SQIB domain, grown on graphene on quartz. The polarizer angle has been rotated by \SI{90}{\degree} in between the two measurements. The positions of the two maxima almost agree well with the case of SQIB platelets on glass, dashed vertical lines. Both, (b) polarized optical reflection microscopy (crossed polarizers, polarizer positions indicated by white cross) as well as the AFM micrograph (c) of platelets on graphene reveal a similar appearance as for orthorhombic phase platelets on glass, see Figure~S4. }
\label{absorbance_graphene}
\end{figure}

For a nominally \SI{30}{nm} thick OMBD grown film on graphene/quartz, deposited at a substrate temperature of \SI{100}{\celsius}, the \hkl(110)-oriented orthorhombic phase forms exclusively. The polarized absorbance spectra in Figure~\ref{absorbance_graphene}(a) show that the two absorbance bands visible in the unpolarized spectra, see Figure~\ref{xrd}(b) red curve, are polarized mutually perpendicular within the plane of the thin film. The complete polarization analysis is shown in the Supporting Information Figure~S9 and confirms that the optical absorbance properties are very similar to the orthorhombic phase platelets on glass, Figures~S4.
Just the peak maxima vary slightly: the UDC peaks at \SI{652}{nm} (\SI{645}{nm} glass sample) and the LDC at \SI{740}{nm} (\SI{737}{nm} glass sample) giving a Davydov splitting energy of \SI{0.23}{eV} (\SI{0.24}{eV} glass sample). Also the morphology determined by polarized optical microscopy as well as by AFM, see Figures~\ref{absorbance_graphene}(b) and (c), is similar. Here the domain size, however, is much smaller, dictated by the domain size of the graphene substrate. The uniformness of the optical image within these domains also suggests an epitaxial alignment. This means that the platelets on graphene are rather single crystalline, mutually rotated domains.\cite{Balzer13a} Platelets on non-templating glass substrates can show a gradual rotation of the in-plane orientation within a single domain.\cite{Balzer17a,Funke21} This is evident from the gradual contrast change within a platelet image through a single polarizer, Figure~\ref{moviestills} (d)-(f), or crossed polarizers, Figure~S4(b).

\subsection{OMBD of SQIB on \ce{KCl} \hkl(001)}
\begin{figure}[h]
\centering
\includegraphics[width=0.48\textwidth]{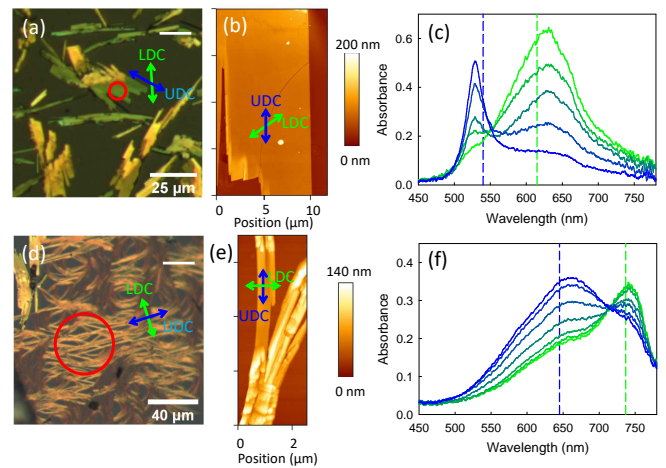}
\caption{Polarized reflection microscopy images (single horizontal polarizer indicated by white horizontal line) of SQIB evaporated onto \ce{KCl}\hkl(001) at a substrate temperature of \SI{120}{\celsius}, forming crystallites (a) and fibers (d). AFM images of a crystallites and fibers, (b) and (e), respectively. Corresponding polarized absorbance spectra, (c) and (f), demonstrate that the aggregates constitute of different polymorphs. The dashed vertical lines mark the spectral positions of the maxima for spincasted SQIB films on glass. For the spectra, the sample has been rotated over \SI{90}{\degree} and \SI{60}{\degree}, respectively, in steps of \SI{15}{\degree}. The red circles in the microscope images mark the position, where the absorbance spectra have been taken. The UDC and LDC directions are depicted in the microscope images by blue and green arrows, respectively. }
\label{KCl_spektren}
\end{figure}

For the thermal deposition of SQIB on freshly cleaved and annealed \ce{KCl} \hkl(001) at elevated substrate temperature of \SI{120}{\celsius}, X-ray diffraction in Figure~\ref{xrd}(b) hints to an excessive formation of the \hkl(011)-oriented monoclinic polymorph with minor regions of the \hkl(110)-oriented orthorhombic phase.
The morphology, however, is quite different from the previously described systems. Optical microscopy and AFM, Figures~\ref{KCl_spektren}(a,b) and also Figures~S10 and S11, show the predominant existence of elongated crystallites with flat surface, but also spaghetti-like fibers appear,  Figures~\ref{KCl_spektren}(d,e). An assignment of the polymorphic phase can only be done by polarized spectro-microscopy owing to its spatially resolving nature.

The flat crystallites appear green to yellow imaged between crossed polarizers with the color impression depending on the orientation of the crystallites. Polarized absorbance spectra as shown in Figure~\ref{KCl_spektren}(c) confirm that the crystallites consist of the monoclinic polymorph. Here, two individual peaks are clearly resolved, while unpolarized absorbance measurements show a broad absorbance band with a flat plateau, Figure~\ref{xrd}(b). The the Davydov splitting energy of the monoclinic phase on \ce{KCl} amounts to \SI{0.37}{eV} which is larger compared to the monoclinic phase formed by postannealing on glass substrates being \SI{0.28}{eV}: UDC at \SI{529}{nm} (\SI{540}{nm} on glass) and LDC at \SI{629}{nm} (\SI{615}{nm} on glass).
The angle $\Delta$ between polarizer orientations for the maxima of the UDC and LDC for the monoclinic polymorph is around \SI{60\pm 6}{\degree}, Figure~S11(c) whereas for the orthorhombic polymorph the angle is $\Delta = \SI{90}{\degree}$, Figure~S4. This value comes close to the expected angle of \SI{56}{\degree} for the case of the monoclinic phase.\cite{Balzer17a}

To a minor extent also spaghetti-like fibers are present, Figure~\ref{KCl_spektren}(d). These fibers can clearly be identified to consist of the orthorhombic polymorph by their absorbance spectrum, Figure~\ref{KCl_spektren}(f). Compared to the platelet spectra on glass, Figure~S4(d), the peaks are broader but have a slightly smaller Davydov splitting energy of \SI{0.2}{eV}: UDC at \SI{662}{nm} (\SI{645}{nm} on glass) and LDC at \SI{743}{nm} (\SI{743}{nm} on glass). The peak broadening might be related to the fact that not a single fiber is measured but the absorbance is averaged over several fibers not fully parallel aligned within the field of view. The polarization angle difference $\Delta$ between maxima of the UDC and LDC for the fibers amounts to \SI{90\pm 3}{\degree}, that is in full agreement with the polarization behavior predicted by molecular exciton theory,\cite{Hestand18} which has already been confirmed
for the \hkl(110)-oriented orthorhombic polymorph platelets on glass, Figure~S5.\cite{Balzer17a,Funke21} Here, the extended fiber-like shape on \ce{KCl}\hkl(001) instead of the platelet shape suggest an epitaxial alignment of the orthorhombic phase with one of the \ce{KCl} high symmetry directions during the growth process.

Due to the micro-sized crystalline texture of the discontinuous thin film, the direction of maximum reflectivity or minimum absorbance can be correlated with the crystallite/fiber orientations.\cite{Balzer13} Polarization analysis plots and histograms are shown in the Supporting Information in Figure~S12 for the maximum reflectivity at $\lambda=\SI{650\pm 5}{nm}$, selected by an interference filter inserted into the microscope. This probes both the LDC of the monoclinic phase crystallites and the UDC of the orthorhombic phase spaghetti-like fibers. The direction of UDC for the monoclinic phase crystallites and LDC for the orthorhombic phase fibers have been extracted from the local polarized transmission spectra, Figure~S11. For the monoclinic phase crystallite the UDC is found to be polarized along the long crystallite axis as depicted by the blue arrows in Figures~\ref{KCl_spektren}(a,b). From the correlation of the LDC direction with the long crystal axis, Figure~S12(b), a specific mean value of polarization direction for maximum reflectivity relative to the long crystallite direction of $|\beta| = \SI{58\pm 4}{\degree}$ is obtained. The direction of the LDC is depicted by green arrows in Figures~\ref{KCl_spektren}(a,b).
This is consistent with analysis of the spatially resolved polarized absorbance spectra, Figure~S11. For the orthorhombic phase spaghetti-like fibers, the UDC is likewise polarized along the long fiber axis as indicated by blue arrows in Figures~\ref{KCl_spektren}(d,e). The green arrows depict the direction of the LDC, being rotated by $\SI{90}{\degree}$ and therefore is oriented along the short fiber axis.

\subsection{OMBD of SQIB on \ce{Au}\hkl(111)/Mica}

\begin{figure}[h]
\centering
\includegraphics[width=0.48\textwidth]{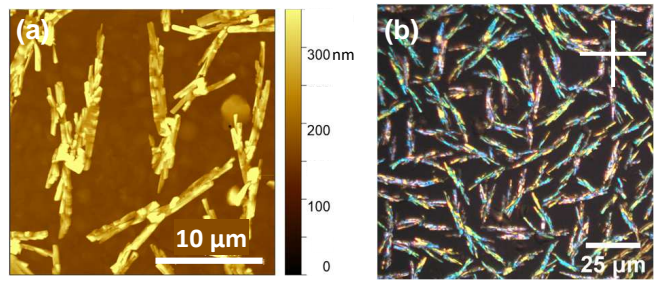}
\caption{(a) AFM image of SQIB on \ce{Au}\hkl(111), nominal thickness of \SI{30}{nm}. A corresponding optical microscope image (reflection, crossed polarizers indicated by horizontal and vertical white line), (b), provides a large scale impression. }
\label{AFM}
\end{figure}

Finally, the vacuum deposition of a \SI{30}{nm} nominal thick SQIB layer on the metallic substrate \ce{Au}\hkl(111)/mica is investigated. A submonolayer coverage of SQIB on silver surfaces showed the formation of a wetting layer.\cite{Luft18} Here, for a nominally \SI{30}{nm} thick film of the monoclinic phase again with the \hkl(011) orientation is formed as deduced from XRD, Figure~\ref{xrd}(a). In Figures~\ref{AFM}(a) and (b), an AFM image as well as a reflection microscopy image using crossed polarizers are shown. The large height of the fibrous elongated crystallites, about ten times higher than the nominal film thickness, hints to dewetting.
The colorful impression of the fibrous crystallites demonstrate the polycrystalline nature with small domain sizes similar to the annealing-induced monoclinic phase on glass substrates. While on glass the fibrous crystallites tend to grow away from a crystallization seed resulting in a dense, flower-like morphology, here separated elongated fibrous aggregates are formed. Similar to the glass sample, also for monoclinic phase on gold the crystalline domain size is too small to be analyzed by the polarized spectro-microscopy setup. Only on \ce{KCl} the monoclinic SQIB phase grows into large enough domains suitable for further optical analysis.

\section{Summary and Conclusions}

\begin{table*}

\begin{tabular}{|c|c|c|c|c|}
\hline
processing & substrate  & temperature & polymorph & orientation \\
OMBD & this work &  substrate &  &  \\
\hline
\includegraphics[width=0.08\textwidth]{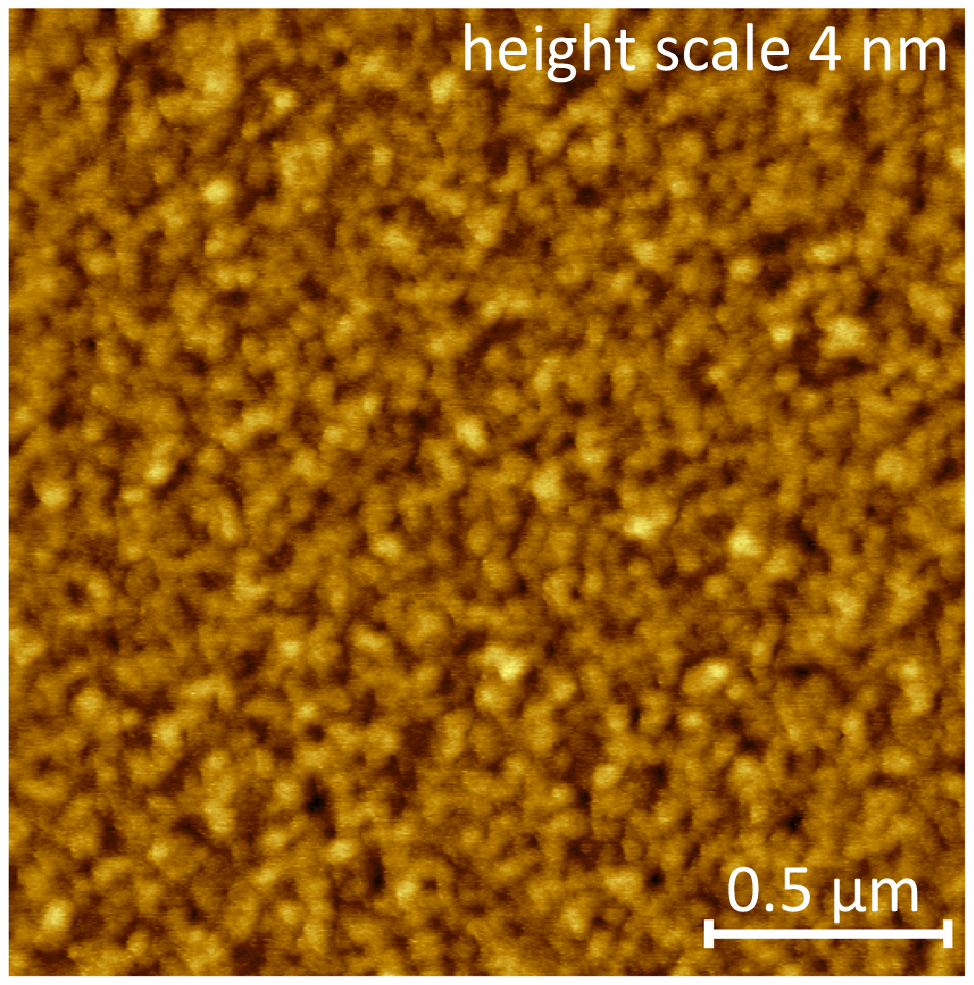} & \ce{SiO2}   & \SI{100}{\celsius} & amorphous & isotropic \\
\includegraphics[width=0.08\textwidth]{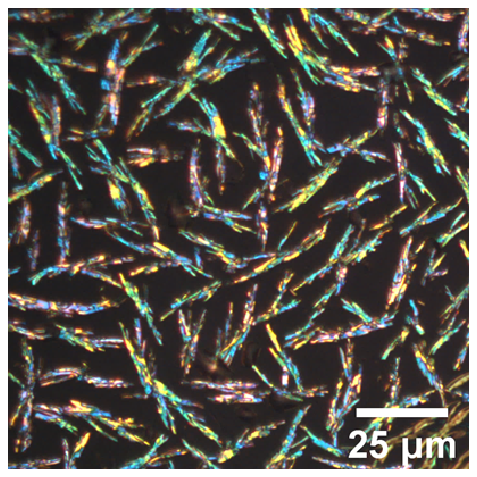} & \ce{Au}\hkl(111)  & \SI{100}{\celsius} & P21/c  & \hkl(011) \\
\includegraphics[width=0.08\textwidth]{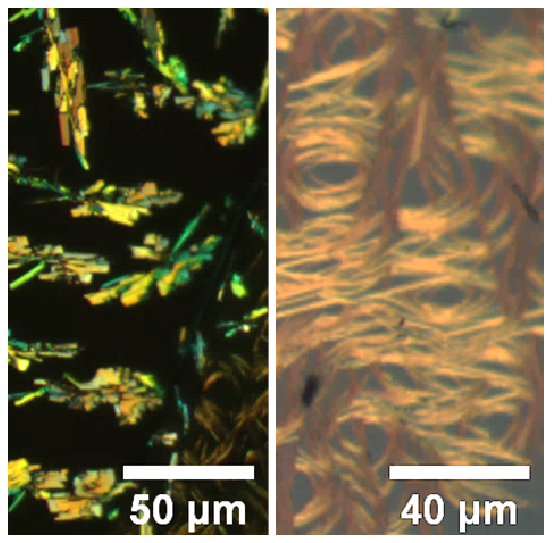} & \ce{KCl}  & \SI{120}{\celsius} & P21/c + Pbcn  & \hkl(011) + \hkl(110) \\
\includegraphics[width=0.08\textwidth]{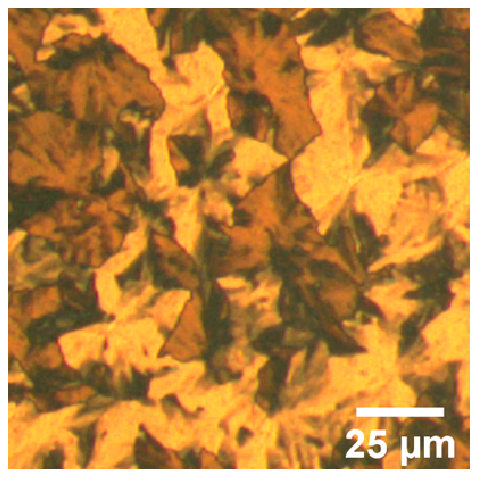} & graphene  & \SI{100}{\celsius} & Pbcn  & \hkl(110) \\
\hline
\hline
processing & substrate  & temperature & polymorph & orientation \\
spincasting & previous work &  postannealing &  &  \\
\hline
\includegraphics[width=0.08\textwidth]{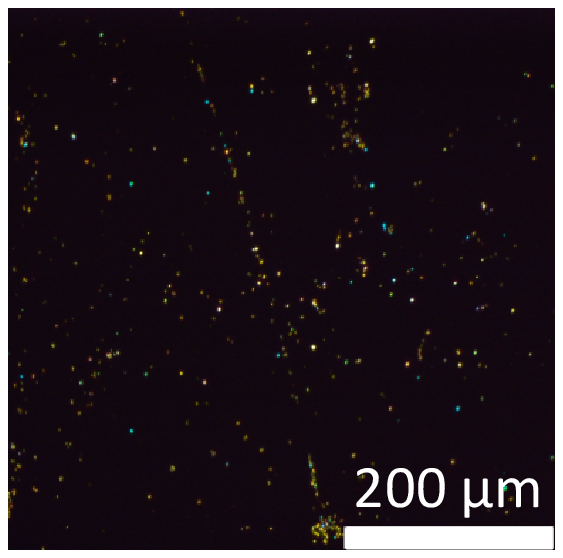} & glass\cite{Balzer17a}   & none  & amorphous & isotropic \\
\includegraphics[width=0.08\textwidth]{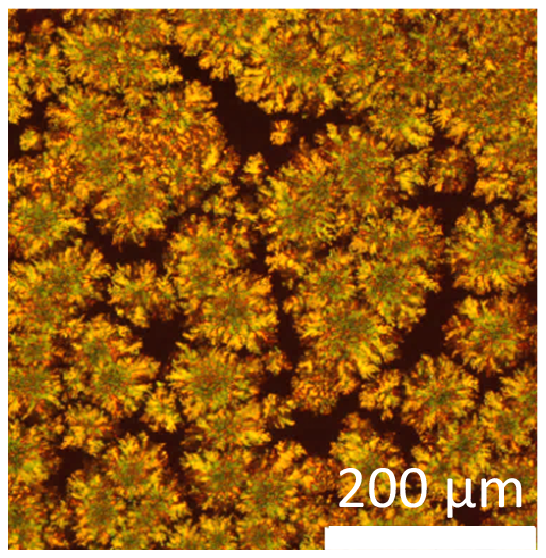} & glass\cite{Balzer17a}   & \SI{90}{\celsius} & P21/c  & \hkl(011) \\
\includegraphics[width=0.08\textwidth]{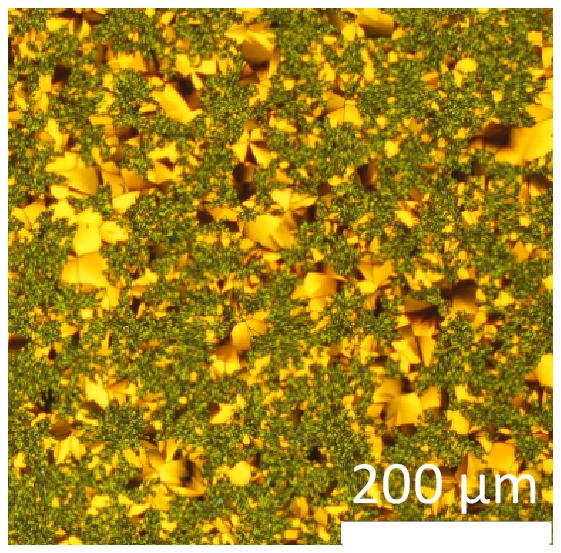} & glass\cite{Balzer17a}   &  \SI{120}{\celsius} & P21/c + Pbcn  & \hkl(011) + \hkl(110) \\
\includegraphics[width=0.08\textwidth]{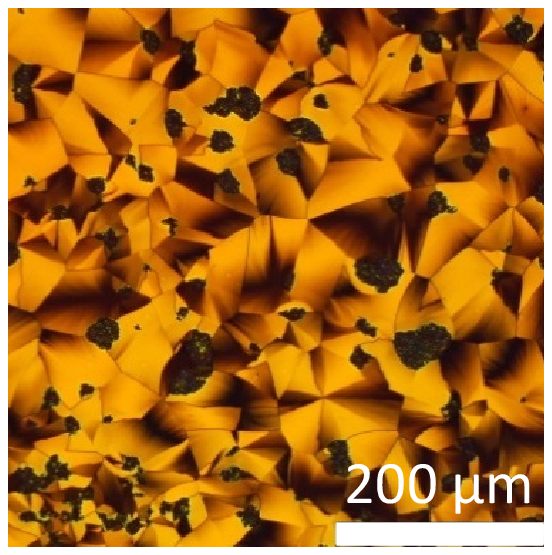} & glass\cite{Balzer17a,Funke21}   & \SI{180}{\celsius} & Pbcn  & \hkl(110) \\
\hline
\hline
\includegraphics[width=0.08\textwidth]{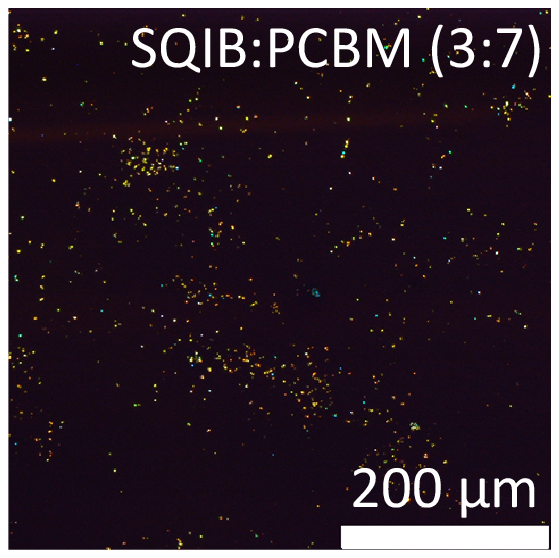} & \ce{MoO3},\cite{Scheunemann17,Scheunemann2019} PEDOT:PSS\cite{Scheunemann2019}   & \SI{60}{\celsius}  & amorphous & isotropic \\
\includegraphics[width=0.08\textwidth]{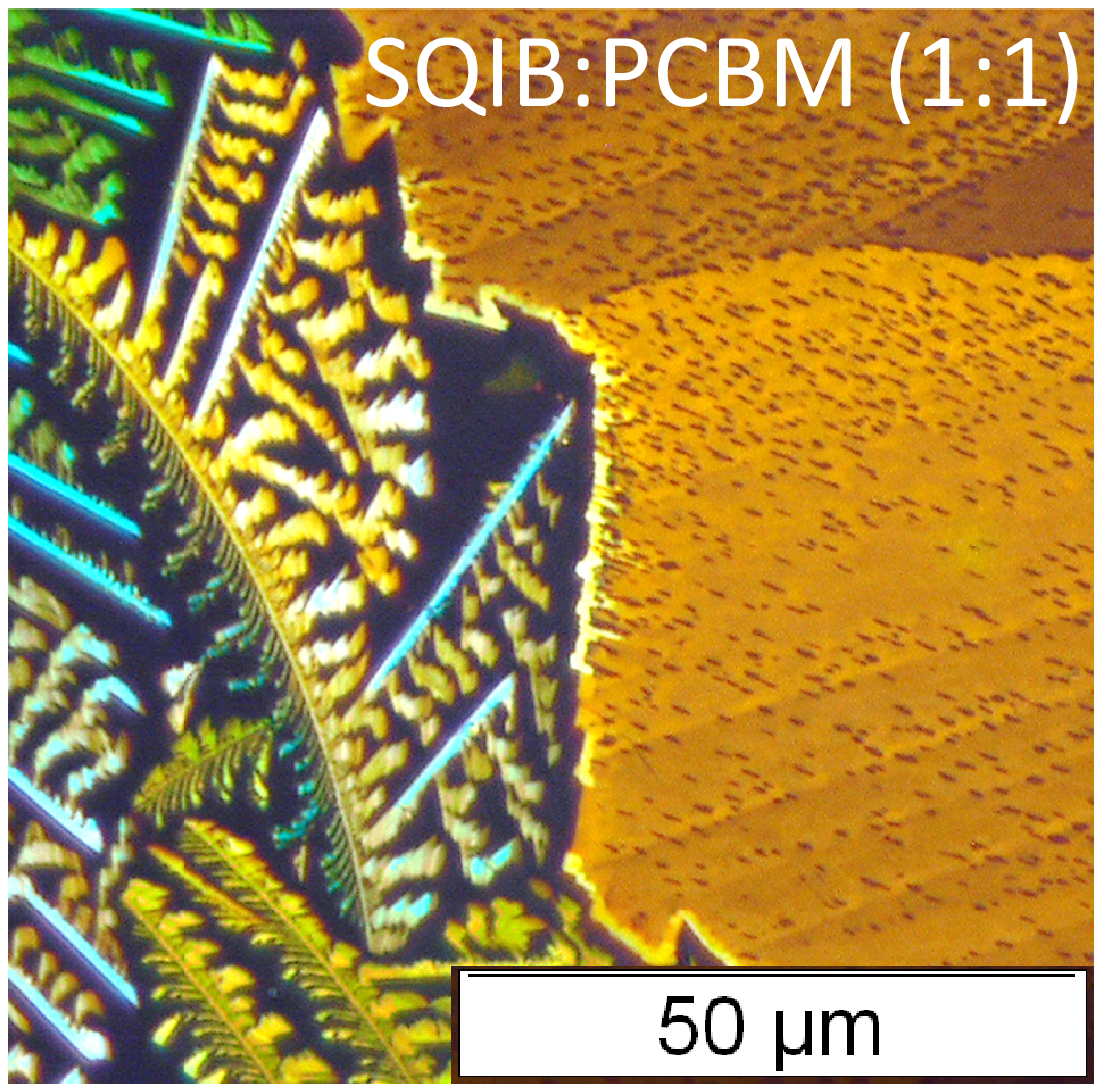} & ITO\cite{Abdullaeva16,Abdullaeva18,Balzer19}   &  \SI{180}{\celsius} & P21/c + Pbcn  & \hkl(011) + \hkl(110) \\
\hline
\end{tabular}

\caption{Summary of polymorph selection by templating of vapor deposited SQIB thin films (this work) and by post-annealing of solution processed thin films (previous works). All images are optical microscopy images between crossed polarizers except for SQIB on \ce{SiO2} which is an AFM image. P21/c = monoclinic polymorph, Pbcn = orthorhombic polymorph. Spincasting SQIB:PCBM blends on \ce{MoO3}, PEDOT:PSS and ITO gives the same results as spincasting on glass, which are not shown.}
\label{summary}

\end{table*}

The anilino squaraine with iso-butyl terminal functionalization (SQIB) is a prototypical donor-type molecular semiconductor suitable for photovoltaic applications that can equifeasible be deposited by spincasting and by thermal vapor deposition. The two known polymorphs of SQIB, a monoclinic and an orthorhombic phase, can be selected by heterogeneous nucleation from the gas phase on dielectric and conductive surfaces -- such as silicon dioxide, potassium chloride, graphene and gold (templating), and by the post-annealing temperature of solution processed thin films, see summary in Table~\ref{summary}. The rapid evaporation of the solvent is suspected to result in a metastable molecular arrangement leaving room for thermal activation of molecular reorganization. This is especially interesting for applications that desire a controlled crystalline microscopic patterning.
For the vacuum deposited thin films at a specific deposition temperature, the polymorph is templated by the growth substrate. For a non-templating substrate, here a \ce{SiO2}-coated wafer, a stable amorphous and isotropic phase is formed, that cannot be recrystallized by subsequent thermal treatment. Since the amorphous phase is favored for light-harvesting applications, which need to tolerate elevated device operation temperatures, vapor deposition appears to be the preferred processing technique.
The three phases of SQIB (amorphous, monoclinic, orthorhombic) can easily be distinguished by their absorbance spectra and their birefringent appearance. The pronounced molecular interactions, characteristic for squaraine compounds due to their D-A-D backbone, are dominating in the bulk phase of both polymorphs. Therefore, the same orientation is obtained for each polymorph on every substrate, so that the molecular stacking direction always runs parallel to the surface. While with that the out-of-plane orientation is fixed, the morphology and in-plane arrangement can to some extend be controlled by epitaxial alignment.

\section{Materials and Methods}

\subsection{Sample Preparation}
2,4-bis[4-(\textit{N,N}-diisobutyl\-amino)-2,6-dihydroxyphenyl]squaraine (SQIB) has been synthesized following our previously published procedure.\cite{Abdullaeva16}
It crystallizes into two different polymorphs: single crystal structures of the monoclinic P21/c (CCDC code 1567209) and of the orthorhombic Pbcn phases (CCDC code 1567104) have been published earlier.\cite{Balzer17a} For deposition via spincoating, a \SI{5}{mg/mL} solution of SQIB in chloroform (Sigma-Aldrich, stabilized with amylene) was prepared. The solution was  spincasted in inert atmosphere at \SI{3000}{rpm}, ramping 3, for \SI{60}{s} (SÜSS MicroTec LabSpin). This was followed by annealing on a preheated hotplate at the indicated surface temperatures for minutes to hours (IKA yellow line). The surface temperature was validated by a thermocouple contact thermometer.

In vacuum (base pressure $p=\SI{1E-7}{mbar}$), samples are prepared by organic molecular beam deposition (OMBD).
A substrate temperature of \SI{100}{\celsius} has been chosen for \ce{SiO2}, graphene/quartz and \ce{Au}\hkl(111)/mica, but for \ce{KCl} substrate temperature was raised to \SI{120}{\celsius}. Likewise, the same nominal thickness of \SI{30}{nm} was deposited for the sake of comparability. Note that the nominal thickness refers to the readout of a Quartz micro-balance and the true film thickness on the sample can deviate, especially for textured thin films.

\subsection{In Situ Optical Characterization}
Temperature-resolved optical microscopy was performed using an optical microscope in reflection geometry in combination with a heatable sample holder. The temperature was controlled using a K-type thermocouple that was placed directly next to the sample.

\subsection{X-Ray Diffraction}
X-ray diffraction (XRD) on thin films were performed in Bragg-Brentano geometry with an automatic divergence slit. A PANalytical X’PertPro MPD diffractometer using Cu-K$\alpha$ radiation ($\lambda = \SI{0.1541}{\angstrom}$) was used, the tube set to \SI{40}{kV} and \SI{40}{mA} with a \SI{10}{mm} beam mask. Samples were rotated in a sample spinner during measurement to eliminate possible effects from preferential in-plane orientation. In addition, a Bruker D8 Discovery diffractometer using Cu-K$\alpha$ radiation was used.

\subsection{Scanning Probe Microscopy}
Morphological characterization took place by atomic force microscopy (AFM, JPK Nano\-wizard). Typically, intermittent contact images were taken with BudgetSensors Tap300-G tips with a force constant of \SI{40}{N/m}, a resonance frequency \SI{300}{kHz}, and a tip radius smaller than \SI{10}{nm}. The AFM was situated on an inverted optical microscope (Nikon Eclipse TE 300) to allow simultaneous optical and morphological characterization.
In addition, an Agilent SPM 5500 system was used operated in tapping mode with MikroMasch cantilevers, resonance frequency \SI{325}{kHz}, spring constant \SI{40}{N/m}.
Samples were imaged under ambient conditions. Gwyddion \cite{Necas12} as well as the software provided by the AFM manufacturers has been used for image analysis.

\subsection{Ellipsometry}
The complex refractive index of isotropic films has been determined by variable angle spectroscopic ellipsometry utilizing a J.A. Woollam rotating analyzer ellipsometer (VASE) with vertical sample stage. Standard ellipsometric data in reflection as well as p-polarized reflection have been recorded with WVASE32 software. Parameters: AOI = \SI{15}{\degree}, \SI{35}{\degree}, \SI{55}{\degree}, AutoSlit = \SI{1}{mm}, wavelength steps = \SI{5}{nm}, wavelength range = \SI{350}{nm} to \SI{1700}{nm}. The ellipsometric data were converted to CompleteEASE (CE) format and analyzed with CE version 6 using Multi Sample Analysis to fit samples with various layer thicknesses simultaneously for decoupling of fit parameters.\cite{Zablocki2020c} Here, two samples of \SI{25}{nm} and \SI{48}{nm} thickness have been analyzed. The layer thickness was determined within the transparent spectral range from \SI{900}{nm} to \SI{1700}{nm} by a Cauchy model. This was then converted to a model free, Kramers-Kronig-consistent isotropic B-spline fit and extended over the full spectral range with \SI{0.1}{eV} resolution (except \SI{660}{nm} to \SI{720}{nm} with \SI{0.05}{eV} resolution).

For the \ce{Si} substrate with native oxide, the database complex refractive index "SI\_JAW" fits well, and the native oxide layer was determined to be \SI{2.02}{nm} thick using "NTVE\_JAW" database complex refractive index.

\subsection{Polarized Optical Characterization}
For the basic polarized optical characterization, a polarization microscope (Leitz DMRME) was used. The projected orientations of the upper and lower Davydov components within the thin films were determined by polarized reflection and transmission microscopy. Illumination took place either by linear polarized white light or by quasi-monochromatic light, selected by bandpass filters (Thorlabs FKB-VIS-10 and VEB Carl Zeiss Jena) with a FWHM of \SI{10}{nm} and \SI{7.5}{nm}, respectively. To determine, spatially resolved, the LDC and UDC directions, the sample was rotated in steps of \SI{5}{\degree} over \SI{360}{\degree} by a computer controlled rotation stage. For each angle a microscope image was taken. The series of images was analysed in ImageJ\cite{Schneider12} by a discrete Fourier transform.\cite{Balzer13,Balzer15}
From the intensity variation $I^{x,y}$ of the pixel at position $(x,y)$, the angle $\phi_{\mathrm{1pol}}\left(x,y\right)$ for the largest reflectivity or transmission is calculated.
To correlate the polarization angle $\phi_{\mathrm{1pol}}(x,y)$ with the crystallite or fiber directions, their local orientation $\theta_{\mathrm{orient}}\left(x,y\right)$ at position $(x,y)$ is determined with help of the structure tensor.\cite{Rezakhaniha11,Balzer15} From this, the angle of maximum reflectivity at a certain wavelength with respect to the long crystallite direction, $\beta = \phi_{\mathrm{1pol}}-\theta_{\mathrm{orient}}$, is determined. Spatially resolved, polarized transmission and reflection spectra were measured in a similar way with a fiber-optics miniature spectrometer (Ocean Optics Maya2000), coupled via a \SI{200}{\micro m} diameter fiber to the camera port of the microscope.

\begin{acknowledgements}
MS thanks the PRO RETINA Stiftung, the DFG (RTG 1885 “Molecular Basis of Sensory Biology”) as well as the Linz Institute of Technology (LIT-2019-7-INC-313 SEAMBIOF) for funding, and is grateful to Prof. em. Jürgen Parisi, University of Oldenburg, for providing access to excellent facilities including technical support at the Energy and Semiconductor Research Laboratory. We are indebted to Matthias Schulz and Arne Lützen, University of Bonn, for providing SQIB powder.
\end{acknowledgements}

%


\begin{thebibliography}{49}%
\makeatletter
\providecommand \@ifxundefined [1]{%
 \@ifx{#1\undefined}
}%
\providecommand \@ifnum [1]{%
 \ifnum #1\expandafter \@firstoftwo
 \else \expandafter \@secondoftwo
 \fi
}%
\providecommand \@ifx [1]{%
 \ifx #1\expandafter \@firstoftwo
 \else \expandafter \@secondoftwo
 \fi
}%
\providecommand \natexlab [1]{#1}%
\providecommand \enquote  [1]{``#1''}%
\providecommand \bibnamefont  [1]{#1}%
\providecommand \bibfnamefont [1]{#1}%
\providecommand \citenamefont [1]{#1}%
\providecommand \href@noop [0]{\@secondoftwo}%
\providecommand \href [0]{\begingroup \@sanitize@url \@href}%
\providecommand \@href[1]{\@@startlink{#1}\@@href}%
\providecommand \@@href[1]{\endgroup#1\@@endlink}%
\providecommand \@sanitize@url [0]{\catcode `\\12\catcode `\$12\catcode
  `\&12\catcode `\#12\catcode `\^12\catcode `\_12\catcode `\%12\relax}%
\providecommand \@@startlink[1]{}%
\providecommand \@@endlink[0]{}%
\providecommand \url  [0]{\begingroup\@sanitize@url \@url }%
\providecommand \@url [1]{\endgroup\@href {#1}{\urlprefix }}%
\providecommand \urlprefix  [0]{URL }%
\providecommand \Eprint [0]{\href }%
\providecommand \doibase [0]{https://doi.org/}%
\providecommand \selectlanguage [0]{\@gobble}%
\providecommand \bibinfo  [0]{\@secondoftwo}%
\providecommand \bibfield  [0]{\@secondoftwo}%
\providecommand \translation [1]{[#1]}%
\providecommand \BibitemOpen [0]{}%
\providecommand \bibitemStop [0]{}%
\providecommand \bibitemNoStop [0]{.\EOS\space}%
\providecommand \EOS [0]{\spacefactor3000\relax}%
\providecommand \BibitemShut  [1]{\csname bibitem#1\endcsname}%
\let\auto@bib@innerbib\@empty
\bibitem [{\citenamefont {Kahr}\ \emph {et~al.}(2010)\citenamefont {Kahr},
  \citenamefont {Freudenthal},\ and\ \citenamefont {Gunn}}]{Kahr2010}%
  \BibitemOpen
  \bibfield  {author} {\bibinfo {author} {\bibfnamefont {B.}~\bibnamefont
  {Kahr}}, \bibinfo {author} {\bibfnamefont {J.}~\bibnamefont {Freudenthal}},\
  and\ \bibinfo {author} {\bibfnamefont {E.}~\bibnamefont {Gunn}},\ }\bibfield
  {title} {\bibinfo {title} {Crystals in light},\ }\href
  {https://doi.org/10.1021/ar900288m} {\bibfield  {journal} {\bibinfo
  {journal} {Acc. Chem. Res.}\ }\textbf {\bibinfo {volume} {43}},\ \bibinfo
  {pages} {684} (\bibinfo {year} {2010})}\BibitemShut {NoStop}%
\bibitem [{\citenamefont {Brewer}\ \emph {et~al.}(2008)\citenamefont {Brewer},
  \citenamefont {Schiek}, \citenamefont {Wallmann},\ and\ \citenamefont
  {Rubahn}}]{Brewer08}%
  \BibitemOpen
  \bibfield  {author} {\bibinfo {author} {\bibfnamefont {J.}~\bibnamefont
  {Brewer}}, \bibinfo {author} {\bibfnamefont {M.}~\bibnamefont {Schiek}},
  \bibinfo {author} {\bibfnamefont {I.}~\bibnamefont {Wallmann}},\ and\
  \bibinfo {author} {\bibfnamefont {H.-G.}\ \bibnamefont {Rubahn}},\ }\bibfield
   {title} {\bibinfo {title} {First order optical nonlinearities $\chi^2$ for
  organic nanofibers from functionalized \emph{para}-phenylenes},\ }\href
  {https://doi.org/10.1016/j.optcom.2008.03.075} {\bibfield  {journal}
  {\bibinfo  {journal} {Opt. Commun.}\ }\textbf {\bibinfo {volume} {281}},\
  \bibinfo {pages} {3892} (\bibinfo {year} {2008})}\BibitemShut {NoStop}%
\bibitem [{\citenamefont {Zablocki}\ \emph
  {et~al.}(2020{\natexlab{a}})\citenamefont {Zablocki}, \citenamefont
  {Arteaga}, \citenamefont {Balzer}, \citenamefont {Hertel}, \citenamefont
  {Holstein}, \citenamefont {Clever}, \citenamefont {Anh{\"a}user},
  \citenamefont {Puttreddy}, \citenamefont {Rissanen}, \citenamefont
  {Meerholz}, \citenamefont {L{\"u}tzen},\ and\ \citenamefont
  {Schiek}}]{Zablocki20}%
  \BibitemOpen
  \bibfield  {author} {\bibinfo {author} {\bibfnamefont {J.}~\bibnamefont
  {Zablocki}}, \bibinfo {author} {\bibfnamefont {O.}~\bibnamefont {Arteaga}},
  \bibinfo {author} {\bibfnamefont {F.}~\bibnamefont {Balzer}}, \bibinfo
  {author} {\bibfnamefont {D.}~\bibnamefont {Hertel}}, \bibinfo {author}
  {\bibfnamefont {J.}~\bibnamefont {Holstein}}, \bibinfo {author}
  {\bibfnamefont {G.}~\bibnamefont {Clever}}, \bibinfo {author} {\bibfnamefont
  {J.}~\bibnamefont {Anh{\"a}user}}, \bibinfo {author} {\bibfnamefont
  {R.}~\bibnamefont {Puttreddy}}, \bibinfo {author} {\bibfnamefont
  {K.}~\bibnamefont {Rissanen}}, \bibinfo {author} {\bibfnamefont
  {K.}~\bibnamefont {Meerholz}}, \bibinfo {author} {\bibfnamefont
  {A.}~\bibnamefont {L{\"u}tzen}},\ and\ \bibinfo {author} {\bibfnamefont
  {M.}~\bibnamefont {Schiek}},\ }\bibfield  {title} {\bibinfo {title}
  {Polymorphic chiral squaraine crystallites in textured thin films},\ }\href
  {https://doi.org/10.1002/chir.23213} {\bibfield  {journal} {\bibinfo
  {journal} {Chirality}\ }\textbf {\bibinfo {volume} {32}},\ \bibinfo {pages}
  {619} (\bibinfo {year} {2020}{\natexlab{a}})}\BibitemShut {NoStop}%
\bibitem [{\citenamefont {Schulz}\ \emph {et~al.}(2018)\citenamefont {Schulz},
  \citenamefont {Zablocki}, \citenamefont {Abdullaeva}, \citenamefont
  {Br{\"u}ck}, \citenamefont {Balzer}, \citenamefont {L{\"u}tzen},
  \citenamefont {Arteaga},\ and\ \citenamefont {Schiek}}]{Schulz18}%
  \BibitemOpen
  \bibfield  {author} {\bibinfo {author} {\bibfnamefont {M.}~\bibnamefont
  {Schulz}}, \bibinfo {author} {\bibfnamefont {J.}~\bibnamefont {Zablocki}},
  \bibinfo {author} {\bibfnamefont {O.~S.}\ \bibnamefont {Abdullaeva}},
  \bibinfo {author} {\bibfnamefont {S.}~\bibnamefont {Br{\"u}ck}}, \bibinfo
  {author} {\bibfnamefont {F.}~\bibnamefont {Balzer}}, \bibinfo {author}
  {\bibfnamefont {A.}~\bibnamefont {L{\"u}tzen}}, \bibinfo {author}
  {\bibfnamefont {O.}~\bibnamefont {Arteaga}},\ and\ \bibinfo {author}
  {\bibfnamefont {M.}~\bibnamefont {Schiek}},\ }\bibfield  {title} {\bibinfo
  {title} {Giant intrinsic circular dichroism of prolinol-derived squaraine
  thin films},\ }\href {https://doi.org/10.1038/s41467-018-04811-7} {\bibfield
  {journal} {\bibinfo  {journal} {Nat. Commun.}\ }\textbf {\bibinfo {volume}
  {9}},\ \bibinfo {pages} {2413} (\bibinfo {year} {2018})}\BibitemShut
  {NoStop}%
\bibitem [{\citenamefont {Llinas}\ and\ \citenamefont
  {Goodman}(2008)}]{Llinas08}%
  \BibitemOpen
  \bibfield  {author} {\bibinfo {author} {\bibfnamefont {A.}~\bibnamefont
  {Llinas}}\ and\ \bibinfo {author} {\bibfnamefont {J.~M.}\ \bibnamefont
  {Goodman}},\ }\bibfield  {title} {\bibinfo {title} {Polymorph control: past,
  present and future},\ }\href {https://doi.org/10.1016/j.drudis.2007.11.006}
  {\bibfield  {journal} {\bibinfo  {journal} {Drug Discov. Today}\ }\textbf
  {\bibinfo {volume} {13}},\ \bibinfo {pages} {198} (\bibinfo {year}
  {2008})}\BibitemShut {NoStop}%
\bibitem [{\citenamefont {Gentili}\ \emph {et~al.}(2019)\citenamefont
  {Gentili}, \citenamefont {Gazzano}, \citenamefont {Melucci}, \citenamefont
  {Jones},\ and\ \citenamefont {Cavallini}}]{Gentili19}%
  \BibitemOpen
  \bibfield  {author} {\bibinfo {author} {\bibfnamefont {D.}~\bibnamefont
  {Gentili}}, \bibinfo {author} {\bibfnamefont {M.}~\bibnamefont {Gazzano}},
  \bibinfo {author} {\bibfnamefont {M.}~\bibnamefont {Melucci}}, \bibinfo
  {author} {\bibfnamefont {D.}~\bibnamefont {Jones}},\ and\ \bibinfo {author}
  {\bibfnamefont {M.}~\bibnamefont {Cavallini}},\ }\bibfield  {title} {\bibinfo
  {title} {Polymorphism as an additional functionality of materials for
  technological applications at surfaces and interfaces},\ }\href
  {https://doi.org/10.1039/c8cs00283e} {\bibfield  {journal} {\bibinfo
  {journal} {Chem. Soc. Rev.}\ }\textbf {\bibinfo {volume} {48}},\ \bibinfo
  {pages} {2502} (\bibinfo {year} {2019})}\BibitemShut {NoStop}%
\bibitem [{\citenamefont {Jones}\ \emph {et~al.}(2016)\citenamefont {Jones},
  \citenamefont {Chattopadhyay}, \citenamefont {Geerts},\ and\ \citenamefont
  {Resel}}]{Jones16}%
  \BibitemOpen
  \bibfield  {author} {\bibinfo {author} {\bibfnamefont {A.~O.~F.}\
  \bibnamefont {Jones}}, \bibinfo {author} {\bibfnamefont {B.}~\bibnamefont
  {Chattopadhyay}}, \bibinfo {author} {\bibfnamefont {Y.~H.}\ \bibnamefont
  {Geerts}},\ and\ \bibinfo {author} {\bibfnamefont {R.}~\bibnamefont
  {Resel}},\ }\bibfield  {title} {\bibinfo {title} {Substrate-induced and
  thin-film phases: Polymorphism of organic materials on surfaces},\ }\href
  {https://doi.org/10.1002/adfm.201503169} {\bibfield  {journal} {\bibinfo
  {journal} {Adv. Funct. Mater.}\ }\textbf {\bibinfo {volume} {26}},\ \bibinfo
  {pages} {2233 } (\bibinfo {year} {2016})}\BibitemShut {NoStop}%
\bibitem [{\citenamefont {Cruz-Cabeza}\ \emph {et~al.}(2015)\citenamefont
  {Cruz-Cabeza}, \citenamefont {Reutzel-Edens},\ and\ \citenamefont
  {Bernstein}}]{CruzCabeza15}%
  \BibitemOpen
  \bibfield  {author} {\bibinfo {author} {\bibfnamefont {A.~J.}\ \bibnamefont
  {Cruz-Cabeza}}, \bibinfo {author} {\bibfnamefont {S.~M.}\ \bibnamefont
  {Reutzel-Edens}},\ and\ \bibinfo {author} {\bibfnamefont {J.}~\bibnamefont
  {Bernstein}},\ }\bibfield  {title} {\bibinfo {title} {Facts and fictions
  about polymorphism},\ }\href {https://doi.org/10.1039/c5cs00227c} {\bibfield
  {journal} {\bibinfo  {journal} {Chem. Soc. Rev.}\ }\textbf {\bibinfo {volume}
  {44}},\ \bibinfo {pages} {8619} (\bibinfo {year} {2015})}\BibitemShut
  {NoStop}%
\bibitem [{\citenamefont {Lee}\ \emph {et~al.}(2011)\citenamefont {Lee},
  \citenamefont {Park}, \citenamefont {Sim}, \citenamefont {Lim}, \citenamefont
  {Kim}, \citenamefont {Hong},\ and\ \citenamefont {Cho}}]{Lee11}%
  \BibitemOpen
  \bibfield  {author} {\bibinfo {author} {\bibfnamefont {W.}~\bibnamefont
  {Lee}}, \bibinfo {author} {\bibfnamefont {J.}~\bibnamefont {Park}}, \bibinfo
  {author} {\bibfnamefont {S.}~\bibnamefont {Sim}}, \bibinfo {author}
  {\bibfnamefont {S.}~\bibnamefont {Lim}}, \bibinfo {author} {\bibfnamefont
  {K.}~\bibnamefont {Kim}}, \bibinfo {author} {\bibfnamefont {B.}~\bibnamefont
  {Hong}},\ and\ \bibinfo {author} {\bibfnamefont {K.}~\bibnamefont {Cho}},\
  }\bibfield  {title} {\bibinfo {title} {Surface-directed molecular assembly of
  pentacene on monolayer graphene for high-performance organic transistors},\
  }\href {https://doi.org/10.1021/ja1097463} {\bibfield  {journal} {\bibinfo
  {journal} {J. Am. Chem. Soc.}\ }\textbf {\bibinfo {volume} {133}},\ \bibinfo
  {pages} {4447 } (\bibinfo {year} {2011})}\BibitemShut {NoStop}%
\bibitem [{\citenamefont {Bischof}\ \emph {et~al.}(2021)\citenamefont
  {Bischof}, \citenamefont {Zeplichal}, \citenamefont {Anhäuser},
  \citenamefont {Kumar}, \citenamefont {Kind}, \citenamefont {Kramer},
  \citenamefont {Bolte}, \citenamefont {Ivlev}, \citenamefont {Terfort},\ and\
  \citenamefont {Witte}}]{Bischof2021}%
  \BibitemOpen
  \bibfield  {author} {\bibinfo {author} {\bibfnamefont {D.}~\bibnamefont
  {Bischof}}, \bibinfo {author} {\bibfnamefont {M.}~\bibnamefont {Zeplichal}},
  \bibinfo {author} {\bibfnamefont {S.}~\bibnamefont {Anhäuser}}, \bibinfo
  {author} {\bibfnamefont {A.}~\bibnamefont {Kumar}}, \bibinfo {author}
  {\bibfnamefont {M.}~\bibnamefont {Kind}}, \bibinfo {author} {\bibfnamefont
  {F.}~\bibnamefont {Kramer}}, \bibinfo {author} {\bibfnamefont
  {M.}~\bibnamefont {Bolte}}, \bibinfo {author} {\bibfnamefont {S.~I.}\
  \bibnamefont {Ivlev}}, \bibinfo {author} {\bibfnamefont {A.}~\bibnamefont
  {Terfort}},\ and\ \bibinfo {author} {\bibfnamefont {G.}~\bibnamefont
  {Witte}},\ }\bibfield  {title} {\bibinfo {title} {Perfluorinated acenes:
  Crystalline phases, polymorph-selective growth, and optoelectronic
  properties},\ }\href {https://doi.org/10.1021/acs.jpcc.1c05985} {\bibfield
  {journal} {\bibinfo  {journal} {J. Phys. Chem. C}\ }\textbf {\bibinfo
  {volume} {125}},\ \bibinfo {pages} {19000} (\bibinfo {year}
  {2021})}\BibitemShut {NoStop}%
\bibitem [{\citenamefont {Blagden}\ \emph {et~al.}(2007)\citenamefont
  {Blagden}, \citenamefont {de~Matas}, \citenamefont {Gavan},\ and\
  \citenamefont {York}}]{Blagden2007}%
  \BibitemOpen
  \bibfield  {author} {\bibinfo {author} {\bibfnamefont {N.}~\bibnamefont
  {Blagden}}, \bibinfo {author} {\bibfnamefont {M.}~\bibnamefont {de~Matas}},
  \bibinfo {author} {\bibfnamefont {P.}~\bibnamefont {Gavan}},\ and\ \bibinfo
  {author} {\bibfnamefont {P.}~\bibnamefont {York}},\ }\bibfield  {title}
  {\bibinfo {title} {Crystal engineering of active pharmaceutical ingredients
  to improve solubility and dissolution rates},\ }\href
  {https://doi.org/10.1016/j.addr.2007.05.011} {\bibfield  {journal} {\bibinfo
  {journal} {Adv. Drug Delivery Rev.}\ }\textbf {\bibinfo {volume} {59}},\
  \bibinfo {pages} {617} (\bibinfo {year} {2007})}\BibitemShut {NoStop}%
\bibitem [{\citenamefont {Hilfiker}\ and\ \citenamefont {von
  Raumer}()}]{Hilfiker18}%
  \BibitemOpen
  \bibinfo {editor} {\bibfnamefont {R.}~\bibnamefont {Hilfiker}}\ and\ \bibinfo
  {editor} {\bibfnamefont {M.}~\bibnamefont {von Raumer}},\ eds.,\ \href
  {https://doi.org/10.1002/9783527697847} {\emph {\bibinfo {title}
  {Polymorphism in the Pharmaceutical Industry}}}\ (\bibinfo  {publisher}
  {Wiley-{VCH} Verlag {GmbH} {\&} Co. {KGaA}})\BibitemShut {NoStop}%
\bibitem [{\citenamefont {Lee}(2014)}]{Lee14}%
  \BibitemOpen
  \bibfield  {author} {\bibinfo {author} {\bibfnamefont {E.~H.}\ \bibnamefont
  {Lee}},\ }\bibfield  {title} {\bibinfo {title} {A practical guide to
  pharmaceutical polymorph screening {\&} selection},\ }\href
  {https://doi.org/10.1016/j.ajps.2014.05.002} {\bibfield  {journal} {\bibinfo
  {journal} {Asian J. Pharm. Sci.}\ }\textbf {\bibinfo {volume} {9}},\ \bibinfo
  {pages} {163} (\bibinfo {year} {2014})}\BibitemShut {NoStop}%
\bibitem [{\citenamefont {Yang}\ \emph {et~al.}(2017)\citenamefont {Yang},
  \citenamefont {Hu}, \citenamefont {Zhu}, \citenamefont {Zhu}, \citenamefont
  {Ward},\ and\ \citenamefont {Kahr}}]{Yang2017}%
  \BibitemOpen
  \bibfield  {author} {\bibinfo {author} {\bibfnamefont {J.}~\bibnamefont
  {Yang}}, \bibinfo {author} {\bibfnamefont {C.~T.}\ \bibnamefont {Hu}},
  \bibinfo {author} {\bibfnamefont {X.}~\bibnamefont {Zhu}}, \bibinfo {author}
  {\bibfnamefont {Q.}~\bibnamefont {Zhu}}, \bibinfo {author} {\bibfnamefont
  {M.~D.}\ \bibnamefont {Ward}},\ and\ \bibinfo {author} {\bibfnamefont
  {B.}~\bibnamefont {Kahr}},\ }\bibfield  {title} {\bibinfo {title} {{DDT}
  polymorphism and the lethality of crystal forms},\ }\href
  {https://doi.org/10.1002/ange.201703028} {\bibfield  {journal} {\bibinfo
  {journal} {Angew. Chem.}\ }\textbf {\bibinfo {volume} {129}},\ \bibinfo
  {pages} {10299} (\bibinfo {year} {2017})}\BibitemShut {NoStop}%
\bibitem [{\citenamefont {Halton}(2008)}]{Halton08}%
  \BibitemOpen
  \bibfield  {author} {\bibinfo {author} {\bibfnamefont {B.}~\bibnamefont
  {Halton}},\ }\bibfield  {title} {\bibinfo {title} {From small rings to big
  things: Xerography, sensors, and the squaraines},\ }\href@noop {} {\bibfield
  {journal} {\bibinfo  {journal} {Chem. New Zealand}\ }\textbf {\bibinfo
  {volume} {72}},\ \bibinfo {pages} {57} (\bibinfo {year} {2008})}\BibitemShut
  {NoStop}%
\bibitem [{\citenamefont {Weiss}(2016)}]{Weiss16}%
  \BibitemOpen
  \bibfield  {author} {\bibinfo {author} {\bibfnamefont {D.~S.}\ \bibnamefont
  {Weiss}},\ }\bibfield  {title} {\bibinfo {title} {The history and development
  of organic photoconductors for electrophotography},\ }\href
  {https://doi.org/10.2352/j.imagingsci.technol.2016.60.3.030505} {\bibfield
  {journal} {\bibinfo  {journal} {J. Imaging Sci. Technol.}\ }\textbf {\bibinfo
  {volume} {60}},\ \bibinfo {pages} {305051} (\bibinfo {year}
  {2016})}\BibitemShut {NoStop}%
\bibitem [{\citenamefont {Chen}\ \emph {et~al.}(2015)\citenamefont {Chen},
  \citenamefont {Sasabe}, \citenamefont {Igarashi}, \citenamefont {Hong},\ and\
  \citenamefont {Kido}}]{Chen15b}%
  \BibitemOpen
  \bibfield  {author} {\bibinfo {author} {\bibfnamefont {G.}~\bibnamefont
  {Chen}}, \bibinfo {author} {\bibfnamefont {H.}~\bibnamefont {Sasabe}},
  \bibinfo {author} {\bibfnamefont {T.}~\bibnamefont {Igarashi}}, \bibinfo
  {author} {\bibfnamefont {Z.}~\bibnamefont {Hong}},\ and\ \bibinfo {author}
  {\bibfnamefont {J.}~\bibnamefont {Kido}},\ }\bibfield  {title} {\bibinfo
  {title} {Squaraine dyes for organic photovoltaic cells},\ }\href
  {https://doi.org/10.1039/c5ta01879j} {\bibfield  {journal} {\bibinfo
  {journal} {J. Mater. Chem.A}\ }\textbf {\bibinfo {volume} {3}},\ \bibinfo
  {pages} {14517} (\bibinfo {year} {2015})}\BibitemShut {NoStop}%
\bibitem [{\citenamefont {Chen}\ \emph {et~al.}(2019)\citenamefont {Chen},
  \citenamefont {Zhu}, \citenamefont {Wu}, \citenamefont {Huang}, \citenamefont
  {Facchetti},\ and\ \citenamefont {Marks}}]{Chen19}%
  \BibitemOpen
  \bibfield  {author} {\bibinfo {author} {\bibfnamefont {Y.}~\bibnamefont
  {Chen}}, \bibinfo {author} {\bibfnamefont {W.}~\bibnamefont {Zhu}}, \bibinfo
  {author} {\bibfnamefont {J.}~\bibnamefont {Wu}}, \bibinfo {author}
  {\bibfnamefont {Y.}~\bibnamefont {Huang}}, \bibinfo {author} {\bibfnamefont
  {A.}~\bibnamefont {Facchetti}},\ and\ \bibinfo {author} {\bibfnamefont
  {T.~J.}\ \bibnamefont {Marks}},\ }\bibfield  {title} {\bibinfo {title}
  {Recent advances in squaraine dyes for bulk-heterojunction organic solar
  cells},\ }\href {https://doi.org/10.1515/oph-2019-0001} {\bibfield  {journal}
  {\bibinfo  {journal} {Org. Photonics Photovolt.}\ }\textbf {\bibinfo {volume}
  {6}},\ \bibinfo {pages} {1} (\bibinfo {year} {2019})}\BibitemShut {NoStop}%
\bibitem [{\citenamefont {Br{\"u}ck}\ \emph {et~al.}(2014)\citenamefont
  {Br{\"u}ck}, \citenamefont {Krause}, \citenamefont {Turrisi}, \citenamefont
  {Beverina}, \citenamefont {Wilken}, \citenamefont {Saak}, \citenamefont
  {L{\"u}tzen}, \citenamefont {Borchert}, \citenamefont {Schiek},\ and\
  \citenamefont {Parisi}}]{Brueck14}%
  \BibitemOpen
  \bibfield  {author} {\bibinfo {author} {\bibfnamefont {S.}~\bibnamefont
  {Br{\"u}ck}}, \bibinfo {author} {\bibfnamefont {C.}~\bibnamefont {Krause}},
  \bibinfo {author} {\bibfnamefont {R.}~\bibnamefont {Turrisi}}, \bibinfo
  {author} {\bibfnamefont {L.}~\bibnamefont {Beverina}}, \bibinfo {author}
  {\bibfnamefont {S.}~\bibnamefont {Wilken}}, \bibinfo {author} {\bibfnamefont
  {W.}~\bibnamefont {Saak}}, \bibinfo {author} {\bibfnamefont {A.}~\bibnamefont
  {L{\"u}tzen}}, \bibinfo {author} {\bibfnamefont {H.}~\bibnamefont
  {Borchert}}, \bibinfo {author} {\bibfnamefont {M.}~\bibnamefont {Schiek}},\
  and\ \bibinfo {author} {\bibfnamefont {J.}~\bibnamefont {Parisi}},\
  }\bibfield  {title} {\bibinfo {title} {Structure--property relationship of
  anilino-squaraines in organic solar cells},\ }\href
  {https://doi.org/10.1039/c3cp54163k} {\bibfield  {journal} {\bibinfo
  {journal} {Phys. Chem. Chem. Phys.}\ }\textbf {\bibinfo {volume} {16}},\
  \bibinfo {pages} {1067} (\bibinfo {year} {2014})}\BibitemShut {NoStop}%
\bibitem [{\citenamefont {Scheunemann}\ \emph {et~al.}(2017)\citenamefont
  {Scheunemann}, \citenamefont {Kolloge}, \citenamefont {Wilken}, \citenamefont
  {Mack}, \citenamefont {Parisi}, \citenamefont {Schulz}, \citenamefont
  {L{\"u}tzen},\ and\ \citenamefont {Schiek}}]{Scheunemann17}%
  \BibitemOpen
  \bibfield  {author} {\bibinfo {author} {\bibfnamefont {D.}~\bibnamefont
  {Scheunemann}}, \bibinfo {author} {\bibfnamefont {O.}~\bibnamefont
  {Kolloge}}, \bibinfo {author} {\bibfnamefont {S.}~\bibnamefont {Wilken}},
  \bibinfo {author} {\bibfnamefont {M.}~\bibnamefont {Mack}}, \bibinfo {author}
  {\bibfnamefont {J.}~\bibnamefont {Parisi}}, \bibinfo {author} {\bibfnamefont
  {M.}~\bibnamefont {Schulz}}, \bibinfo {author} {\bibfnamefont
  {A.}~\bibnamefont {L{\"u}tzen}},\ and\ \bibinfo {author} {\bibfnamefont
  {M.}~\bibnamefont {Schiek}},\ }\bibfield  {title} {\bibinfo {title}
  {Revealing the recombination dynamics in squaraine-based bulk heterojunction
  solar cells},\ }\href {https://doi.org/10.1063/1.4996080} {\bibfield
  {journal} {\bibinfo  {journal} {Appl. Phys. Lett.}\ }\textbf {\bibinfo
  {volume} {111}},\ \bibinfo {pages} {183502} (\bibinfo {year}
  {2017})}\BibitemShut {NoStop}%
\bibitem [{\citenamefont {Scheunemann}\ \emph {et~al.}(2019)\citenamefont
  {Scheunemann}, \citenamefont {Wilken}, \citenamefont {Sandberg},
  \citenamefont {{\"O}sterbacka},\ and\ \citenamefont
  {Schiek}}]{Scheunemann2019}%
  \BibitemOpen
  \bibfield  {author} {\bibinfo {author} {\bibfnamefont {D.}~\bibnamefont
  {Scheunemann}}, \bibinfo {author} {\bibfnamefont {S.}~\bibnamefont {Wilken}},
  \bibinfo {author} {\bibfnamefont {O.~J.}\ \bibnamefont {Sandberg}}, \bibinfo
  {author} {\bibfnamefont {R.}~\bibnamefont {{\"O}sterbacka}},\ and\ \bibinfo
  {author} {\bibfnamefont {M.}~\bibnamefont {Schiek}},\ }\bibfield  {title}
  {\bibinfo {title} {Effect of imbalanced charge transport on the interplay of
  surface and bulk recombination in organic solar cells},\ }\href
  {https://doi.org/10.1103/physrevapplied.11.054090} {\bibfield  {journal}
  {\bibinfo  {journal} {Phys. Rev. Applied}\ }\textbf {\bibinfo {volume}
  {11}},\ \bibinfo {pages} {054090} (\bibinfo {year} {2019})}\BibitemShut
  {NoStop}%
\bibitem [{\citenamefont {Chen}\ \emph {et~al.}(2018)\citenamefont {Chen},
  \citenamefont {Ling}, \citenamefont {Wei}, \citenamefont {Zhang},
  \citenamefont {Hong}, \citenamefont {Sasabe},\ and\ \citenamefont
  {Kido}}]{Chen2018}%
  \BibitemOpen
  \bibfield  {author} {\bibinfo {author} {\bibfnamefont {G.}~\bibnamefont
  {Chen}}, \bibinfo {author} {\bibfnamefont {Z.}~\bibnamefont {Ling}}, \bibinfo
  {author} {\bibfnamefont {B.}~\bibnamefont {Wei}}, \bibinfo {author}
  {\bibfnamefont {J.}~\bibnamefont {Zhang}}, \bibinfo {author} {\bibfnamefont
  {Z.}~\bibnamefont {Hong}}, \bibinfo {author} {\bibfnamefont {H.}~\bibnamefont
  {Sasabe}},\ and\ \bibinfo {author} {\bibfnamefont {J.}~\bibnamefont {Kido}},\
  }\bibfield  {title} {\bibinfo {title} {Comparison of the solution and
  vacuum-processed squaraine:fullerene small-molecule bulk heterojunction solar
  cells},\ }\bibfield  {journal} {\bibinfo  {journal} {Front. Chem.}\ }\textbf
  {\bibinfo {volume} {6}},\ \href {https://doi.org/10.3389/fchem.2018.00412}
  {10.3389/fchem.2018.00412} (\bibinfo {year} {2018})\BibitemShut {NoStop}%
\bibitem [{\citenamefont {Schulz}\ \emph {et~al.}(2019)\citenamefont {Schulz},
  \citenamefont {Balzer}, \citenamefont {Scheunemann}, \citenamefont {Arteaga},
  \citenamefont {L{\"u}tzen}, \citenamefont {Meskers},\ and\ \citenamefont
  {Schiek}}]{Schulz19}%
  \BibitemOpen
  \bibfield  {author} {\bibinfo {author} {\bibfnamefont {M.}~\bibnamefont
  {Schulz}}, \bibinfo {author} {\bibfnamefont {F.}~\bibnamefont {Balzer}},
  \bibinfo {author} {\bibfnamefont {D.}~\bibnamefont {Scheunemann}}, \bibinfo
  {author} {\bibfnamefont {O.}~\bibnamefont {Arteaga}}, \bibinfo {author}
  {\bibfnamefont {A.}~\bibnamefont {L{\"u}tzen}}, \bibinfo {author}
  {\bibfnamefont {S.}~\bibnamefont {Meskers}},\ and\ \bibinfo {author}
  {\bibfnamefont {M.}~\bibnamefont {Schiek}},\ }\bibfield  {title} {\bibinfo
  {title} {Chiral excitonic organic photodiodes for direct detection of
  circular polarized light},\ }\href {https://doi.org/10.1002/adfm.201900684}
  {\bibfield  {journal} {\bibinfo  {journal} {Adv. Funct. Mater.}\ }\textbf
  {\bibinfo {volume} {29}},\ \bibinfo {pages} {1900684} (\bibinfo {year}
  {2019})}\BibitemShut {NoStop}%
\bibitem [{\citenamefont {Abdullaeva}\ \emph {et~al.}(2019)\citenamefont
  {Abdullaeva}, \citenamefont {Balzer}, \citenamefont {Schulz}, \citenamefont
  {Parisi}, \citenamefont {L{\"u}tzen}, \citenamefont {Dedek},\ and\
  \citenamefont {Schiek}}]{Abdullaeva18}%
  \BibitemOpen
  \bibfield  {author} {\bibinfo {author} {\bibfnamefont {O.~S.}\ \bibnamefont
  {Abdullaeva}}, \bibinfo {author} {\bibfnamefont {F.}~\bibnamefont {Balzer}},
  \bibinfo {author} {\bibfnamefont {M.}~\bibnamefont {Schulz}}, \bibinfo
  {author} {\bibfnamefont {J.}~\bibnamefont {Parisi}}, \bibinfo {author}
  {\bibfnamefont {A.}~\bibnamefont {L{\"u}tzen}}, \bibinfo {author}
  {\bibfnamefont {K.}~\bibnamefont {Dedek}},\ and\ \bibinfo {author}
  {\bibfnamefont {M.}~\bibnamefont {Schiek}},\ }\bibfield  {title} {\bibinfo
  {title} {Organic photovoltaic sensors for photocapacitive stimulation of
  voltage-gated ion channels in neuroblastoma cells},\ }\href
  {https://doi.org/10.1002/adfm.201805177} {\bibfield  {journal} {\bibinfo
  {journal} {Adv. Funct. Mater.}\ }\textbf {\bibinfo {volume} {29}},\ \bibinfo
  {pages} {1805177} (\bibinfo {year} {2019})}\BibitemShut {NoStop}%
\bibitem [{\citenamefont {Balzer}\ \emph {et~al.}(2020)\citenamefont {Balzer},
  \citenamefont {Abdullaeva}, \citenamefont {Maderitsch}, \citenamefont
  {Schulz}, \citenamefont {L{\"u}tzen},\ and\ \citenamefont
  {Schiek}}]{Balzer19}%
  \BibitemOpen
  \bibfield  {author} {\bibinfo {author} {\bibfnamefont {F.}~\bibnamefont
  {Balzer}}, \bibinfo {author} {\bibfnamefont {O.~S.}\ \bibnamefont
  {Abdullaeva}}, \bibinfo {author} {\bibfnamefont {A.}~\bibnamefont
  {Maderitsch}}, \bibinfo {author} {\bibfnamefont {M.}~\bibnamefont {Schulz}},
  \bibinfo {author} {\bibfnamefont {A.}~\bibnamefont {L{\"u}tzen}},\ and\
  \bibinfo {author} {\bibfnamefont {M.}~\bibnamefont {Schiek}},\ }\bibfield
  {title} {\bibinfo {title} {Nanoscale polarization-resolved surface
  photovoltage of a pleochroic squaraine thin film},\ }\href
  {https://doi.org/10.1002/pssb.201900570} {\bibfield  {journal} {\bibinfo
  {journal} {Phys. Status Solidi B}\ }\textbf {\bibinfo {volume} {257}},\
  \bibinfo {pages} {1900570} (\bibinfo {year} {2020})}\BibitemShut {NoStop}%
\bibitem [{\citenamefont {Momma}\ and\ \citenamefont {Izumi}(2011)}]{Momma11}%
  \BibitemOpen
  \bibfield  {author} {\bibinfo {author} {\bibfnamefont {K.}~\bibnamefont
  {Momma}}\ and\ \bibinfo {author} {\bibfnamefont {F.}~\bibnamefont {Izumi}},\
  }\bibfield  {title} {\bibinfo {title} {{VESTA 3} for three-dimensional
  visualization of crystal, volumetric and morphology data},\ }\href
  {https://doi.org/10.1107/S0021889811038970} {\bibfield  {journal} {\bibinfo
  {journal} {J. Appl. Crystallogr.}\ }\textbf {\bibinfo {volume} {44}},\
  \bibinfo {pages} {1272} (\bibinfo {year} {2011})}\BibitemShut {NoStop}%
\bibitem [{\citenamefont {Balzer}\ \emph {et~al.}(2017)\citenamefont {Balzer},
  \citenamefont {Kollmann}, \citenamefont {Schulz}, \citenamefont
  {Schnakenburg}, \citenamefont {L{\"u}tzen}, \citenamefont {Schmidtmann},
  \citenamefont {Lienau}, \citenamefont {Silies},\ and\ \citenamefont
  {Schiek}}]{Balzer17a}%
  \BibitemOpen
  \bibfield  {author} {\bibinfo {author} {\bibfnamefont {F.}~\bibnamefont
  {Balzer}}, \bibinfo {author} {\bibfnamefont {H.}~\bibnamefont {Kollmann}},
  \bibinfo {author} {\bibfnamefont {M.}~\bibnamefont {Schulz}}, \bibinfo
  {author} {\bibfnamefont {G.}~\bibnamefont {Schnakenburg}}, \bibinfo {author}
  {\bibfnamefont {A.}~\bibnamefont {L{\"u}tzen}}, \bibinfo {author}
  {\bibfnamefont {M.}~\bibnamefont {Schmidtmann}}, \bibinfo {author}
  {\bibfnamefont {C.}~\bibnamefont {Lienau}}, \bibinfo {author} {\bibfnamefont
  {M.}~\bibnamefont {Silies}},\ and\ \bibinfo {author} {\bibfnamefont
  {M.}~\bibnamefont {Schiek}},\ }\bibfield  {title} {\bibinfo {title}
  {Spotlight on excitonic coupling in polymorphic and textured anilino
  squaraine thin films},\ }\href {https://doi.org/10.1021/acs.cgd.7b01131}
  {\bibfield  {journal} {\bibinfo  {journal} {Cryst. Growth Des.}\ }\textbf
  {\bibinfo {volume} {17}},\ \bibinfo {pages} {6455} (\bibinfo {year}
  {2017})}\BibitemShut {NoStop}%
\bibitem [{\citenamefont {Viterisi}\ \emph {et~al.}(2014)\citenamefont
  {Viterisi}, \citenamefont {Montcada}, \citenamefont {Kumar}, \citenamefont
  {Gispert-Guirado}, \citenamefont {Martin}, \citenamefont {Escudero},\ and\
  \citenamefont {Palomares}}]{Viterisi2014}%
  \BibitemOpen
  \bibfield  {author} {\bibinfo {author} {\bibfnamefont {A.}~\bibnamefont
  {Viterisi}}, \bibinfo {author} {\bibfnamefont {N.~F.}\ \bibnamefont
  {Montcada}}, \bibinfo {author} {\bibfnamefont {C.~V.}\ \bibnamefont {Kumar}},
  \bibinfo {author} {\bibfnamefont {F.}~\bibnamefont {Gispert-Guirado}},
  \bibinfo {author} {\bibfnamefont {E.}~\bibnamefont {Martin}}, \bibinfo
  {author} {\bibfnamefont {E.}~\bibnamefont {Escudero}},\ and\ \bibinfo
  {author} {\bibfnamefont {E.}~\bibnamefont {Palomares}},\ }\bibfield  {title}
  {\bibinfo {title} {Unambiguous determination of molecular packing in
  crystalline donor domains of small molecule solution processed solar cell
  devices using routine x-ray diffraction techniques},\ }\href
  {https://doi.org/10.1039/c3ta13116e} {\bibfield  {journal} {\bibinfo
  {journal} {J. Mater. Chem.A}\ }\textbf {\bibinfo {volume} {2}},\ \bibinfo
  {pages} {3536} (\bibinfo {year} {2014})}\BibitemShut {NoStop}%
\bibitem [{\citenamefont {Davydov}(1964)}]{Davydov64}%
  \BibitemOpen
  \bibfield  {author} {\bibinfo {author} {\bibfnamefont {A.~S.}\ \bibnamefont
  {Davydov}},\ }\bibfield  {title} {\bibinfo {title} {The theory of molecular
  excitons},\ }\href@noop {} {\bibfield  {journal} {\bibinfo  {journal} {Phys.
  Usp.}\ }\textbf {\bibinfo {volume} {7}},\ \bibinfo {pages} {145} (\bibinfo
  {year} {1964})}\BibitemShut {NoStop}%
\bibitem [{\citenamefont {Kasha}\ \emph {et~al.}(1965)\citenamefont {Kasha},
  \citenamefont {Rawls},\ and\ \citenamefont {{El-Bayoumi}}}]{Kasha65}%
  \BibitemOpen
  \bibfield  {author} {\bibinfo {author} {\bibfnamefont {M.}~\bibnamefont
  {Kasha}}, \bibinfo {author} {\bibfnamefont {H.}~\bibnamefont {Rawls}},\ and\
  \bibinfo {author} {\bibfnamefont {M.}~\bibnamefont {{El-Bayoumi}}},\
  }\bibfield  {title} {\bibinfo {title} {The exciton model in molecular
  spectroscopy},\ }\href {https://doi.org/10.1351/pac196511030371} {\bibfield
  {journal} {\bibinfo  {journal} {Pure Appl. Chem.}\ }\textbf {\bibinfo
  {volume} {11}},\ \bibinfo {pages} {371} (\bibinfo {year} {1965})}\BibitemShut
  {NoStop}%
\bibitem [{\citenamefont {Hestand}\ and\ \citenamefont
  {Spano}(2018)}]{Hestand18}%
  \BibitemOpen
  \bibfield  {author} {\bibinfo {author} {\bibfnamefont {N.~J.}\ \bibnamefont
  {Hestand}}\ and\ \bibinfo {author} {\bibfnamefont {F.~C.}\ \bibnamefont
  {Spano}},\ }\bibfield  {title} {\bibinfo {title} {Expanded theory of {H}- and
  {J}-molecular aggregates: The effect of vibronic coupling and intermolecular
  charge transfer},\ }\href {https://doi.org/10.1021/acs.chemrev.7b00581}
  {\bibfield  {journal} {\bibinfo  {journal} {Chem. Rev.}\ }\textbf {\bibinfo
  {volume} {118}},\ \bibinfo {pages} {7069} (\bibinfo {year}
  {2018})}\BibitemShut {NoStop}%
\bibitem [{\citenamefont {Breuer}\ \emph {et~al.}(2012)\citenamefont {Breuer},
  \citenamefont {Celik}, \citenamefont {Jakob}, \citenamefont {Tonner},\ and\
  \citenamefont {Witte}}]{Breuer2012}%
  \BibitemOpen
  \bibfield  {author} {\bibinfo {author} {\bibfnamefont {T.}~\bibnamefont
  {Breuer}}, \bibinfo {author} {\bibfnamefont {M.~A.}\ \bibnamefont {Celik}},
  \bibinfo {author} {\bibfnamefont {P.}~\bibnamefont {Jakob}}, \bibinfo
  {author} {\bibfnamefont {R.}~\bibnamefont {Tonner}},\ and\ \bibinfo {author}
  {\bibfnamefont {G.}~\bibnamefont {Witte}},\ }\bibfield  {title} {\bibinfo
  {title} {Vibrational davydov splittings and collective mode polarizations in
  oriented organic semiconductor crystals},\ }\href
  {https://doi.org/10.1021/jp304080g} {\bibfield  {journal} {\bibinfo
  {journal} {J. Phys. Chem. C}\ }\textbf {\bibinfo {volume} {116}},\ \bibinfo
  {pages} {14491} (\bibinfo {year} {2012})}\BibitemShut {NoStop}%
\bibitem [{\citenamefont {Funke}\ \emph {et~al.}(2021)\citenamefont {Funke},
  \citenamefont {Duwe}, \citenamefont {Balzer}, \citenamefont {Thiesen},
  \citenamefont {Hingerl},\ and\ \citenamefont {Schiek}}]{Funke21}%
  \BibitemOpen
  \bibfield  {author} {\bibinfo {author} {\bibfnamefont {S.}~\bibnamefont
  {Funke}}, \bibinfo {author} {\bibfnamefont {M.}~\bibnamefont {Duwe}},
  \bibinfo {author} {\bibfnamefont {F.}~\bibnamefont {Balzer}}, \bibinfo
  {author} {\bibfnamefont {P.~H.}\ \bibnamefont {Thiesen}}, \bibinfo {author}
  {\bibfnamefont {K.}~\bibnamefont {Hingerl}},\ and\ \bibinfo {author}
  {\bibfnamefont {M.}~\bibnamefont {Schiek}},\ }\bibfield  {title} {\bibinfo
  {title} {Determining the dielectric tensor of microtextured organic thin
  films by imaging mueller matrix ellipsometry},\ }\href
  {https://doi.org/10.1021/acs.jpclett.1c00317} {\bibfield  {journal} {\bibinfo
   {journal} {J. Phys. Chem. Lett.}\ }\textbf {\bibinfo {volume} {12}},\
  \bibinfo {pages} {3053} (\bibinfo {year} {2021})}\BibitemShut {NoStop}%
\bibitem [{\citenamefont {Freese}\ \emph {et~al.}(2018)\citenamefont {Freese},
  \citenamefont {L{\"a}ssing}, \citenamefont {Jakob}, \citenamefont {Schulz},
  \citenamefont {L{\"u}tzen}, \citenamefont {Schiek},\ and\ \citenamefont
  {Nilius}}]{Freese18}%
  \BibitemOpen
  \bibfield  {author} {\bibinfo {author} {\bibfnamefont {S.}~\bibnamefont
  {Freese}}, \bibinfo {author} {\bibfnamefont {P.}~\bibnamefont {L{\"a}ssing}},
  \bibinfo {author} {\bibfnamefont {R.}~\bibnamefont {Jakob}}, \bibinfo
  {author} {\bibfnamefont {M.}~\bibnamefont {Schulz}}, \bibinfo {author}
  {\bibfnamefont {A.}~\bibnamefont {L{\"u}tzen}}, \bibinfo {author}
  {\bibfnamefont {M.}~\bibnamefont {Schiek}},\ and\ \bibinfo {author}
  {\bibfnamefont {N.}~\bibnamefont {Nilius}},\ }\bibfield  {title} {\bibinfo
  {title} {Photoluminescence of squaraine thin films: Spatial homogeneity and
  temperature dependence},\ }\href {https://doi.org/10.1002/pssb.201800450}
  {\bibfield  {journal} {\bibinfo  {journal} {Phys. Status Solidi B}\ }\textbf
  {\bibinfo {volume} {256}},\ \bibinfo {pages} {1800450} (\bibinfo {year}
  {2018})}\BibitemShut {NoStop}%
\bibitem [{\citenamefont {Tian}\ \emph {et~al.}(2002)\citenamefont {Tian},
  \citenamefont {Furuki}, \citenamefont {Iwasa}, \citenamefont {Sato},
  \citenamefont {Pu},\ and\ \citenamefont {Tatsuura}}]{Tian02}%
  \BibitemOpen
  \bibfield  {author} {\bibinfo {author} {\bibfnamefont {M.}~\bibnamefont
  {Tian}}, \bibinfo {author} {\bibfnamefont {M.}~\bibnamefont {Furuki}},
  \bibinfo {author} {\bibfnamefont {I.}~\bibnamefont {Iwasa}}, \bibinfo
  {author} {\bibfnamefont {Y.}~\bibnamefont {Sato}}, \bibinfo {author}
  {\bibfnamefont {L.}~\bibnamefont {Pu}},\ and\ \bibinfo {author}
  {\bibfnamefont {S.}~\bibnamefont {Tatsuura}},\ }\bibfield  {title} {\bibinfo
  {title} {Search for squaraine derivatives that can be sublimed without
  thermal decomposition},\ }\href {https://doi.org/10.1021/jp013698r}
  {\bibfield  {journal} {\bibinfo  {journal} {J. Phys. Chem. B}\ }\textbf
  {\bibinfo {volume} {106}},\ \bibinfo {pages} {4370} (\bibinfo {year}
  {2002})}\BibitemShut {NoStop}%
\bibitem [{\citenamefont {Luft}\ \emph {et~al.}(2018)\citenamefont {Luft},
  \citenamefont {Gro{\ss}}, \citenamefont {Schulz}, \citenamefont {L{\"u}tzen},
  \citenamefont {Schiek},\ and\ \citenamefont {Nilius}}]{Luft18}%
  \BibitemOpen
  \bibfield  {author} {\bibinfo {author} {\bibfnamefont {M.}~\bibnamefont
  {Luft}}, \bibinfo {author} {\bibfnamefont {B.}~\bibnamefont {Gro{\ss}}},
  \bibinfo {author} {\bibfnamefont {M.}~\bibnamefont {Schulz}}, \bibinfo
  {author} {\bibfnamefont {A.}~\bibnamefont {L{\"u}tzen}}, \bibinfo {author}
  {\bibfnamefont {M.}~\bibnamefont {Schiek}},\ and\ \bibinfo {author}
  {\bibfnamefont {N.}~\bibnamefont {Nilius}},\ }\bibfield  {title} {\bibinfo
  {title} {Adsorption of squaraine molecules to {Au}(111) and {Ag}(001)
  surfaces},\ }\href {https://doi.org/10.1063/1.5017826} {\bibfield  {journal}
  {\bibinfo  {journal} {J. Chem. Phys.}\ }\textbf {\bibinfo {volume} {148}},\
  \bibinfo {pages} {074702} (\bibinfo {year} {2018})}\BibitemShut {NoStop}%
\bibitem [{\citenamefont {Cranston}\ and\ \citenamefont
  {Lessard}(2021)}]{Cranston2021}%
  \BibitemOpen
  \bibfield  {author} {\bibinfo {author} {\bibfnamefont {R.~R.}\ \bibnamefont
  {Cranston}}\ and\ \bibinfo {author} {\bibfnamefont {B.~H.}\ \bibnamefont
  {Lessard}},\ }\bibfield  {title} {\bibinfo {title} {Metal phthalocyanines:
  thin-film formation, microstructure, and physical properties},\ }\href
  {https://doi.org/10.1039/d1ra03853b} {\bibfield  {journal} {\bibinfo
  {journal} {{RSC} Adv.}\ }\textbf {\bibinfo {volume} {11}},\ \bibinfo {pages}
  {21716} (\bibinfo {year} {2021})}\BibitemShut {NoStop}%
\bibitem [{\citenamefont {Zablocki}\ \emph
  {et~al.}(2020{\natexlab{b}})\citenamefont {Zablocki}, \citenamefont {Schulz},
  \citenamefont {Schnakenburg}, \citenamefont {Beverina}, \citenamefont
  {Warzanowski}, \citenamefont {Revelli}, \citenamefont {Gr{\"u}ninger},
  \citenamefont {Balzer}, \citenamefont {Meerholz}, \citenamefont
  {L{\"u}tzen},\ and\ \citenamefont {Schiek}}]{Zablocki2020c}%
  \BibitemOpen
  \bibfield  {author} {\bibinfo {author} {\bibfnamefont {J.}~\bibnamefont
  {Zablocki}}, \bibinfo {author} {\bibfnamefont {M.}~\bibnamefont {Schulz}},
  \bibinfo {author} {\bibfnamefont {G.}~\bibnamefont {Schnakenburg}}, \bibinfo
  {author} {\bibfnamefont {L.}~\bibnamefont {Beverina}}, \bibinfo {author}
  {\bibfnamefont {P.}~\bibnamefont {Warzanowski}}, \bibinfo {author}
  {\bibfnamefont {A.}~\bibnamefont {Revelli}}, \bibinfo {author} {\bibfnamefont
  {M.}~\bibnamefont {Gr{\"u}ninger}}, \bibinfo {author} {\bibfnamefont
  {F.}~\bibnamefont {Balzer}}, \bibinfo {author} {\bibfnamefont
  {K.}~\bibnamefont {Meerholz}}, \bibinfo {author} {\bibfnamefont
  {A.}~\bibnamefont {L{\"u}tzen}},\ and\ \bibinfo {author} {\bibfnamefont
  {M.}~\bibnamefont {Schiek}},\ }\bibfield  {title} {\bibinfo {title}
  {Structure and dielectric properties of anisotropic n-alkyl anilino squaraine
  thin films},\ }\href {https://doi.org/10.1021/acs.jpcc.0c07498} {\bibfield
  {journal} {\bibinfo  {journal} {J. Phys. Chem. C}\ }\textbf {\bibinfo
  {volume} {124}},\ \bibinfo {pages} {22721} (\bibinfo {year}
  {2020}{\natexlab{b}})}\BibitemShut {NoStop}%
\bibitem [{\citenamefont {Larouche}\ and\ \citenamefont
  {Martinu}(2008)}]{Larouche08}%
  \BibitemOpen
  \bibfield  {author} {\bibinfo {author} {\bibfnamefont {S.}~\bibnamefont
  {Larouche}}\ and\ \bibinfo {author} {\bibfnamefont {L.}~\bibnamefont
  {Martinu}},\ }\bibfield  {title} {\bibinfo {title} {Openfilters: open-source
  software for the design, optimization, and synthesis of optical filters},\
  }\href {https://doi.org/10.1364/ao.47.00c219} {\bibfield  {journal} {\bibinfo
   {journal} {Appl. Opt.}\ }\textbf {\bibinfo {volume} {47}},\ \bibinfo {pages}
  {C219} (\bibinfo {year} {2008})}\BibitemShut {NoStop}%
\bibitem [{\citenamefont {Chen}\ \emph {et~al.}(2012)\citenamefont {Chen},
  \citenamefont {Yokoyama}, \citenamefont {Sasabe}, \citenamefont {Hong},
  \citenamefont {Yang},\ and\ \citenamefont {Kido}}]{Chen12b}%
  \BibitemOpen
  \bibfield  {author} {\bibinfo {author} {\bibfnamefont {G.}~\bibnamefont
  {Chen}}, \bibinfo {author} {\bibfnamefont {D.}~\bibnamefont {Yokoyama}},
  \bibinfo {author} {\bibfnamefont {H.}~\bibnamefont {Sasabe}}, \bibinfo
  {author} {\bibfnamefont {Z.}~\bibnamefont {Hong}}, \bibinfo {author}
  {\bibfnamefont {Y.}~\bibnamefont {Yang}},\ and\ \bibinfo {author}
  {\bibfnamefont {J.}~\bibnamefont {Kido}},\ }\bibfield  {title} {\bibinfo
  {title} {Optical and electrical properties of a squaraine dye in photovoltaic
  cells},\ }\href {https://doi.org/10.1063/1.4747623} {\bibfield  {journal}
  {\bibinfo  {journal} {Appl. Phys. Lett.}\ }\textbf {\bibinfo {volume}
  {101}},\ \bibinfo {pages} {083904} (\bibinfo {year} {2012})}\BibitemShut
  {NoStop}%
\bibitem [{\citenamefont {Vezie}\ \emph {et~al.}(2016)\citenamefont {Vezie},
  \citenamefont {Few}, \citenamefont {Meager}, \citenamefont {Pieridou},
  \citenamefont {D{\"o}rling}, \citenamefont {Ashraf}, \citenamefont
  {Go{\~{n}}i}, \citenamefont {Bronstein}, \citenamefont {McCulloch},
  \citenamefont {Hayes}, \citenamefont {Campoy-Quiles},\ and\ \citenamefont
  {Nelson}}]{Vezie16}%
  \BibitemOpen
  \bibfield  {author} {\bibinfo {author} {\bibfnamefont {M.~S.}\ \bibnamefont
  {Vezie}}, \bibinfo {author} {\bibfnamefont {S.}~\bibnamefont {Few}}, \bibinfo
  {author} {\bibfnamefont {I.}~\bibnamefont {Meager}}, \bibinfo {author}
  {\bibfnamefont {G.}~\bibnamefont {Pieridou}}, \bibinfo {author}
  {\bibfnamefont {B.}~\bibnamefont {D{\"o}rling}}, \bibinfo {author}
  {\bibfnamefont {R.~S.}\ \bibnamefont {Ashraf}}, \bibinfo {author}
  {\bibfnamefont {A.~R.}\ \bibnamefont {Go{\~{n}}i}}, \bibinfo {author}
  {\bibfnamefont {H.}~\bibnamefont {Bronstein}}, \bibinfo {author}
  {\bibfnamefont {I.}~\bibnamefont {McCulloch}}, \bibinfo {author}
  {\bibfnamefont {S.~C.}\ \bibnamefont {Hayes}}, \bibinfo {author}
  {\bibfnamefont {M.}~\bibnamefont {Campoy-Quiles}},\ and\ \bibinfo {author}
  {\bibfnamefont {J.}~\bibnamefont {Nelson}},\ }\bibfield  {title} {\bibinfo
  {title} {Exploring the origin of high optical absorption in conjugated
  polymers},\ }\href {https://doi.org/10.1038/nmat4645} {\bibfield  {journal}
  {\bibinfo  {journal} {Nature Mater.}\ }\textbf {\bibinfo {volume} {15}},\
  \bibinfo {eid} {10.1038/nmat4645} (\bibinfo {year} {2016})}\BibitemShut
  {NoStop}%
\bibitem [{\citenamefont {Zheng}\ \emph {et~al.}(2020)\citenamefont {Zheng},
  \citenamefont {Mark}, \citenamefont {Wiegand}, \citenamefont {Diaz},
  \citenamefont {Cody}, \citenamefont {Spano}, \citenamefont {McCamant},\ and\
  \citenamefont {Collison}}]{Zheng2020}%
  \BibitemOpen
  \bibfield  {author} {\bibinfo {author} {\bibfnamefont {C.}~\bibnamefont
  {Zheng}}, \bibinfo {author} {\bibfnamefont {M.~F.}\ \bibnamefont {Mark}},
  \bibinfo {author} {\bibfnamefont {T.}~\bibnamefont {Wiegand}}, \bibinfo
  {author} {\bibfnamefont {S.~A.}\ \bibnamefont {Diaz}}, \bibinfo {author}
  {\bibfnamefont {J.}~\bibnamefont {Cody}}, \bibinfo {author} {\bibfnamefont
  {F.~C.}\ \bibnamefont {Spano}}, \bibinfo {author} {\bibfnamefont {D.~W.}\
  \bibnamefont {McCamant}},\ and\ \bibinfo {author} {\bibfnamefont {C.~J.}\
  \bibnamefont {Collison}},\ }\bibfield  {title} {\bibinfo {title} {Measurement
  and theoretical interpretation of exciton diffusion as a function of
  intermolecular separation for squaraines targeted for bulk heterojunction
  solar cells},\ }\href {https://doi.org/10.1021/acs.jpcc.9b11816} {\bibfield
  {journal} {\bibinfo  {journal} {J. Phys. Chem. C}\ }\textbf {\bibinfo
  {volume} {124}},\ \bibinfo {pages} {4032} (\bibinfo {year}
  {2020})}\BibitemShut {NoStop}%
\bibitem [{\citenamefont {Balzer}\ \emph
  {et~al.}(2014{\natexlab{a}})\citenamefont {Balzer}, \citenamefont
  {Henrichsen}, \citenamefont {Klarskov}, \citenamefont {Booth}, \citenamefont
  {Sun}, \citenamefont {Parisi}, \citenamefont {Schiek},\ and\ \citenamefont
  {B{\o}ggild}}]{Balzer13a}%
  \BibitemOpen
  \bibfield  {author} {\bibinfo {author} {\bibfnamefont {F.}~\bibnamefont
  {Balzer}}, \bibinfo {author} {\bibfnamefont {H.}~\bibnamefont {Henrichsen}},
  \bibinfo {author} {\bibfnamefont {M.}~\bibnamefont {Klarskov}}, \bibinfo
  {author} {\bibfnamefont {T.}~\bibnamefont {Booth}}, \bibinfo {author}
  {\bibfnamefont {R.}~\bibnamefont {Sun}}, \bibinfo {author} {\bibfnamefont
  {J.}~\bibnamefont {Parisi}}, \bibinfo {author} {\bibfnamefont
  {M.}~\bibnamefont {Schiek}},\ and\ \bibinfo {author} {\bibfnamefont
  {P.}~\bibnamefont {B{\o}ggild}},\ }\bibfield  {title} {\bibinfo {title}
  {Directed self-assembled crystalline oligomer domains on graphene and
  graphite},\ }\href {https://doi.org/10.1088/0957-4484/25/3/035602} {\bibfield
   {journal} {\bibinfo  {journal} {Nanotechnology}\ }\textbf {\bibinfo {volume}
  {25}},\ \bibinfo {pages} {035602} (\bibinfo {year}
  {2014}{\natexlab{a}})}\BibitemShut {NoStop}%
\bibitem [{\citenamefont {Balzer}\ \emph
  {et~al.}(2014{\natexlab{b}})\citenamefont {Balzer}, \citenamefont {Schiek},
  \citenamefont {Osadnik}, \citenamefont {Wallmann}, \citenamefont {Parisi},
  \citenamefont {Rubahn},\ and\ \citenamefont {L{\"u}tzen}}]{Balzer13}%
  \BibitemOpen
  \bibfield  {author} {\bibinfo {author} {\bibfnamefont {F.}~\bibnamefont
  {Balzer}}, \bibinfo {author} {\bibfnamefont {M.}~\bibnamefont {Schiek}},
  \bibinfo {author} {\bibfnamefont {A.}~\bibnamefont {Osadnik}}, \bibinfo
  {author} {\bibfnamefont {I.}~\bibnamefont {Wallmann}}, \bibinfo {author}
  {\bibfnamefont {J.}~\bibnamefont {Parisi}}, \bibinfo {author} {\bibfnamefont
  {H.-G.}\ \bibnamefont {Rubahn}},\ and\ \bibinfo {author} {\bibfnamefont
  {A.}~\bibnamefont {L{\"u}tzen}},\ }\bibfield  {title} {\bibinfo {title}
  {Substrate steered crystallization of naphthyl end-capped oligothiophenes
  into nanowires: The influence of methoxy-functionalization},\ }\href
  {https://doi.org/10.1039/C3CP53881H} {\bibfield  {journal} {\bibinfo
  {journal} {Phys. Chem. Chem. Phys.}\ }\textbf {\bibinfo {volume} {16}},\
  \bibinfo {pages} {5747} (\bibinfo {year} {2014}{\natexlab{b}})}\BibitemShut
  {NoStop}%
\bibitem [{\citenamefont {Abdullaeva}\ \emph {et~al.}(2016)\citenamefont
  {Abdullaeva}, \citenamefont {Schulz}, \citenamefont {Balzer}, \citenamefont
  {Parisi}, \citenamefont {L{\"u}tzen}, \citenamefont {Dedek},\ and\
  \citenamefont {Schiek}}]{Abdullaeva16}%
  \BibitemOpen
  \bibfield  {author} {\bibinfo {author} {\bibfnamefont {O.~S.}\ \bibnamefont
  {Abdullaeva}}, \bibinfo {author} {\bibfnamefont {M.}~\bibnamefont {Schulz}},
  \bibinfo {author} {\bibfnamefont {F.}~\bibnamefont {Balzer}}, \bibinfo
  {author} {\bibfnamefont {J.}~\bibnamefont {Parisi}}, \bibinfo {author}
  {\bibfnamefont {A.}~\bibnamefont {L{\"u}tzen}}, \bibinfo {author}
  {\bibfnamefont {K.}~\bibnamefont {Dedek}},\ and\ \bibinfo {author}
  {\bibfnamefont {M.}~\bibnamefont {Schiek}},\ }\bibfield  {title} {\bibinfo
  {title} {Photoelectrical stimulation of neuronal cells by an organic
  semiconductor-electrolyte interface},\ }\href
  {https://doi.org/10.1021/acs.langmuir.6b02085} {\bibfield  {journal}
  {\bibinfo  {journal} {Langmuir}\ }\textbf {\bibinfo {volume} {32}},\ \bibinfo
  {pages} {8533} (\bibinfo {year} {2016})}\BibitemShut {NoStop}%
\bibitem [{\citenamefont {Necas}\ and\ \citenamefont
  {Klapetek}(2012)}]{Necas12}%
  \BibitemOpen
  \bibfield  {author} {\bibinfo {author} {\bibfnamefont {D.}~\bibnamefont
  {Necas}}\ and\ \bibinfo {author} {\bibfnamefont {P.}~\bibnamefont
  {Klapetek}},\ }\bibfield  {title} {\bibinfo {title} {Gwyddion: an open-source
  software for {SPM} data analysis},\ }\href
  {https://doi.org/10.2478/s11534-011-0096-2} {\bibfield  {journal} {\bibinfo
  {journal} {Cent. Eur. J. Phys.}\ }\textbf {\bibinfo {volume} {10}},\ \bibinfo
  {pages} {181} (\bibinfo {year} {2012})}\BibitemShut {NoStop}%
\bibitem [{\citenamefont {Schneider}\ \emph {et~al.}(2012)\citenamefont
  {Schneider}, \citenamefont {Rasband},\ and\ \citenamefont
  {Eliceiri}}]{Schneider12}%
  \BibitemOpen
  \bibfield  {author} {\bibinfo {author} {\bibfnamefont {C.~A.}\ \bibnamefont
  {Schneider}}, \bibinfo {author} {\bibfnamefont {W.~S.}\ \bibnamefont
  {Rasband}},\ and\ \bibinfo {author} {\bibfnamefont {K.~W.}\ \bibnamefont
  {Eliceiri}},\ }\bibfield  {title} {\bibinfo {title} {{NIH} image to {ImageJ}:
  25 years of image analysis},\ }\href {https://doi.org/10.1038/nmeth.2089}
  {\bibfield  {journal} {\bibinfo  {journal} {Nat. Methods}\ }\textbf {\bibinfo
  {volume} {9}},\ \bibinfo {pages} {671} (\bibinfo {year} {2012})}\BibitemShut
  {NoStop}%
\bibitem [{\citenamefont {Balzer}\ and\ \citenamefont
  {Schiek}(2015)}]{Balzer15}%
  \BibitemOpen
  \bibfield  {author} {\bibinfo {author} {\bibfnamefont {F.}~\bibnamefont
  {Balzer}}\ and\ \bibinfo {author} {\bibfnamefont {M.}~\bibnamefont
  {Schiek}},\ }\bibfield  {title} {\bibinfo {title} {Automated polarized
  microscopy analysis of fluorescent and birefringent nano- and microfibers},\
  }in\ \href {https://doi.org/10.1007/978-3-319-19410-3_7} {\emph {\bibinfo
  {booktitle} {Bottom-Up Self-Organization in Supramolecular Soft Matter}}},\
  \bibinfo {series} {Springer Series in Materials Science}, Vol.\ \bibinfo
  {volume} {217},\ \bibinfo {editor} {edited by\ \bibinfo {editor}
  {\bibfnamefont {S.~C.}\ \bibnamefont {M{\"u}ller}}\ and\ \bibinfo {editor}
  {\bibfnamefont {J.}~\bibnamefont {Parisi}}}\ (\bibinfo  {publisher}
  {Springer},\ \bibinfo {address} {Berlin},\ \bibinfo {year} {2015})\
  Chap.~\bibinfo {chapter} {7}, pp.\ \bibinfo {pages} {151--176}\BibitemShut
  {NoStop}%
\bibitem [{\citenamefont {Rezakhaniha}\ \emph {et~al.}(2012)\citenamefont
  {Rezakhaniha}, \citenamefont {Agianniotis}, \citenamefont {Schrauwen},
  \citenamefont {Griffa}, \citenamefont {Sage}, \citenamefont {Bouten},
  \citenamefont {{van de Vosse}}, \citenamefont {Unser},\ and\ \citenamefont
  {Stergiopulos}}]{Rezakhaniha11}%
  \BibitemOpen
  \bibfield  {author} {\bibinfo {author} {\bibfnamefont {R.}~\bibnamefont
  {Rezakhaniha}}, \bibinfo {author} {\bibfnamefont {A.}~\bibnamefont
  {Agianniotis}}, \bibinfo {author} {\bibfnamefont {J.}~\bibnamefont
  {Schrauwen}}, \bibinfo {author} {\bibfnamefont {A.}~\bibnamefont {Griffa}},
  \bibinfo {author} {\bibfnamefont {D.}~\bibnamefont {Sage}}, \bibinfo {author}
  {\bibfnamefont {C.}~\bibnamefont {Bouten}}, \bibinfo {author} {\bibfnamefont
  {F.}~\bibnamefont {{van de Vosse}}}, \bibinfo {author} {\bibfnamefont
  {M.}~\bibnamefont {Unser}},\ and\ \bibinfo {author} {\bibfnamefont
  {N.}~\bibnamefont {Stergiopulos}},\ }\bibfield  {title} {\bibinfo {title}
  {Experimental investigation of collagen waviness and orientation in the
  arterial adventitia using confocal laser scanning microscopy},\ }\href
  {https://doi.org/10.1007/s10237-011-0325-z} {\bibfield  {journal} {\bibinfo
  {journal} {Biomech. Model. Mechanobiol.}\ }\textbf {\bibinfo {volume} {11}},\
  \bibinfo {pages} {461} (\bibinfo {year} {2012})}\BibitemShut {NoStop}%
\end{thebibliography}

%
%

%

\end{document}


\title{Supporting Information \\ Template and Temperature Controlled Polymorph Formation in Squaraine Thin Films}

\author{Frank Balzer}
\affiliation{SDU Centre for Photonics Engineering, University of Southern Denmark, Alsion 2, DK-6400 S{\o}nderborg, Denmark}

\author{Tobias Breuer}
\affiliation{Department of Physics, Philipps University of Marburg, D-35032 Marburg, Germany}

\author{Gregor Witte}
\affiliation{Department of Physics, Philipps University of Marburg, D-35032 Marburg, Germany}

\author{Manuela Schiek}
\email{manuela.schiek@jku.at}
\affiliation{Institute of Physics, University of Oldenburg, D-26111 Oldenburg, Germany}
\affiliation{Center for Surface- and Nanoanalytics \& Linz Institute for Organic Solar Cells, Johannes Kepler University, A-4040 Linz, Austria}

\makeatletter
\renewcommand{\thefigure}{S\@arabic\c@figure}
\renewcommand{\thetable}{S\@arabic\c@table}

\maketitle



\begin{figure}[h]
\centering
\includegraphics[width=0.45\textwidth]{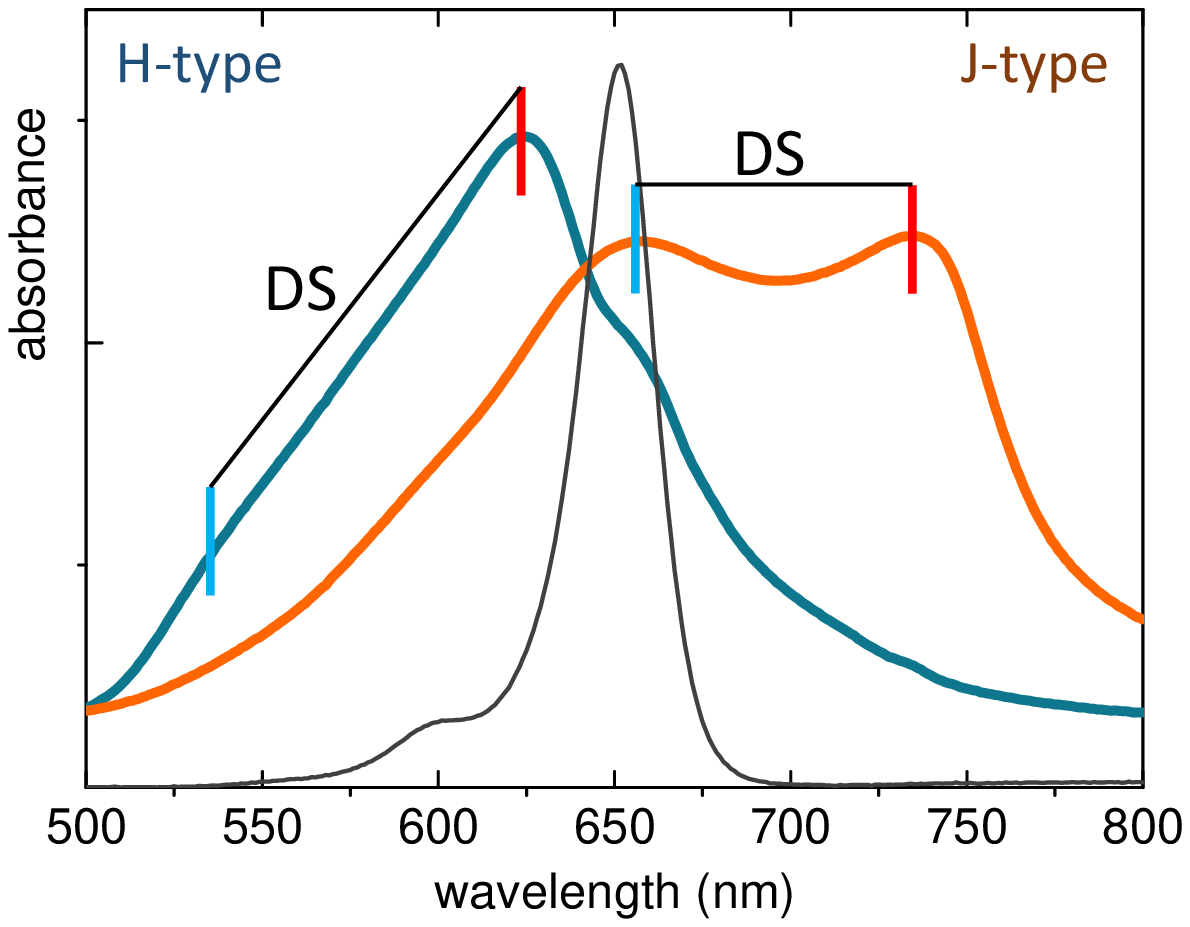}
\caption{Unpolarized absorbance spectra of SQIB: monomer dissolved in chloroform (grey line), monoclinic polymorph (blue line), and orthorhombic polymorph (orange line).}
\label{SQIB-UVVIS}
\end{figure}
The monomer absorbance peaks sharply at \SI{650}{nm} and has a vibronic progression at roughly \SI{600}{nm}. Both the monoclinic and the orthorhombic polymorph are supported on glass with an out-of-plane alignment with the \hkl(011) and \hkl(110) plane parallel to the substrate. The absorbance spectra in normal incidence transmission show a pronounced projected Davydov splitting (DS) as published earlier in reference \citenum{Balzer17a}. The upper (UDC) and lower (LDC) Davydov components are indicated by blue and red bars, respectively. The DS originates from the coupling of translationally inequivalent molecules within the unit cell. One would expect the DS to be centered around the monomer absorbance, resulting in a blue-shifted UDC and a red-shifted LDC, which obviously is not the case here. Instead, both Davydov components are blue-shifted in case of the monoclinic polymorph while they are both red-shifted for the orthorhombic polymorph. This is caused by pronounced coupling between translationally equivalent molecules, that are molecular stacks within the three-dimensional lattice.\cite{Hestand18} In case of the monoclinic polymorph this results in an overall H-type aggregate including the H-like vibronic signatures, which is a relative increase of the first vibronic (0-1) progression. The DS splitting energy between the projected (0-1) UDC and LDC transitions indicated in the graph amounts to \SI{0.24}{eV}. The orthorhombic polymorph can by classified as an overall J-type aggregate with suppressed vibronic progressions. Here, the DS splitting energy amounts to \SI{0.28}{eV} of the indicated (0-0) UDC and LDC transitions.

\clearpage

\begin{figure}[h]
\centering
\includegraphics[width=0.45\textwidth]{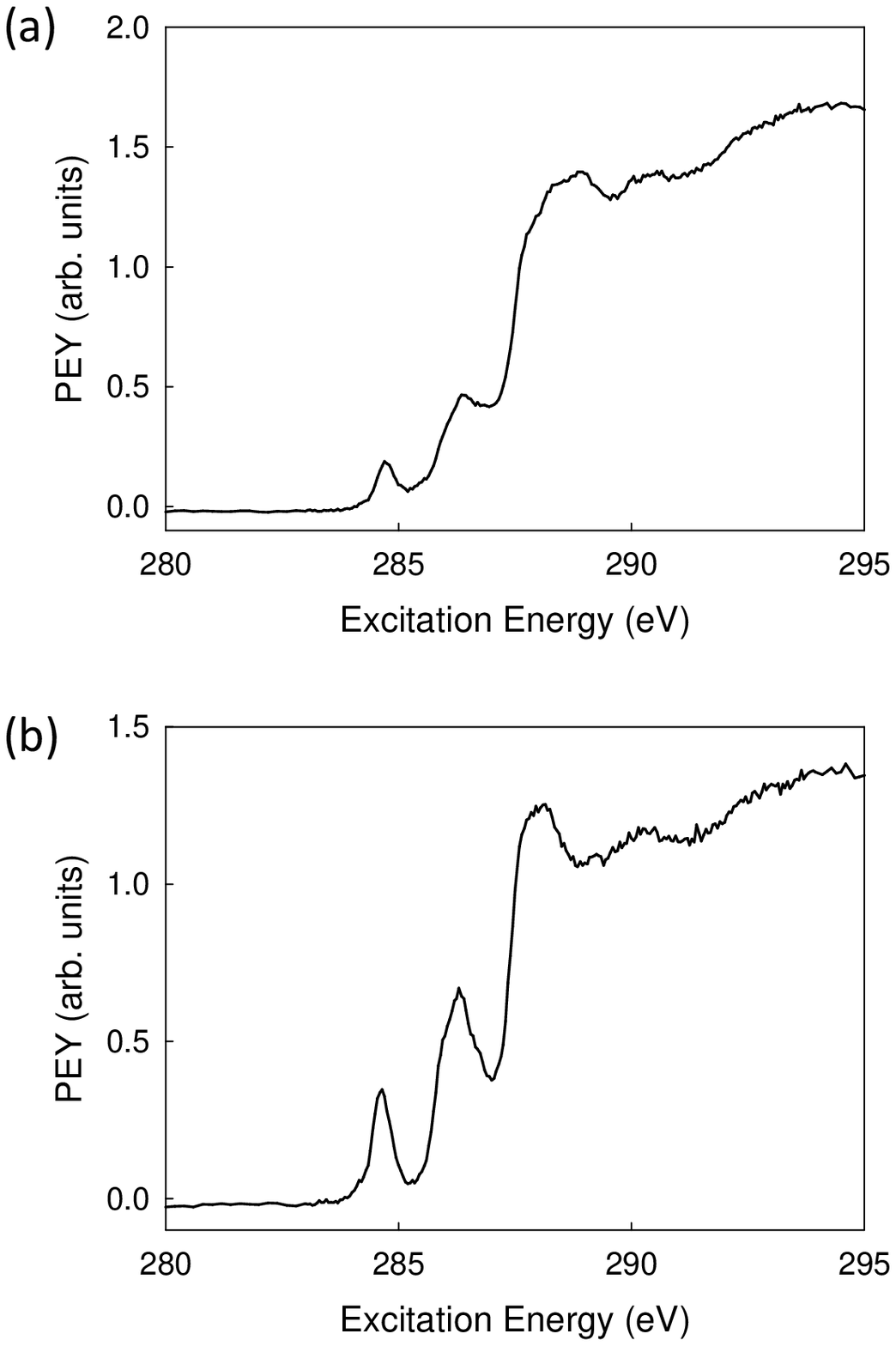}
\caption{Comparison of C1s-NEXAFS data for an evaporated SQIB film on \ce{SiO2} (\ce{Si}-wafer covered with native oxide) with nominal thickness of \SI{30}{nm} (a) and of SQIB powder (b) shows intact sublimation and resublimation of SQIB. No decomposition of the molecules occurs. }
\label{nexafs}
\end{figure}
NEXAFS data of SQIB powder and a vapor deposited thin film on silicon show the same features, and thus consist of the same molecular compound, documenting that the SQIB molecules are stable during thermal vapor deposition.

\clearpage

\begin{figure}[h]
\centering
\includegraphics[width=0.55\textwidth]{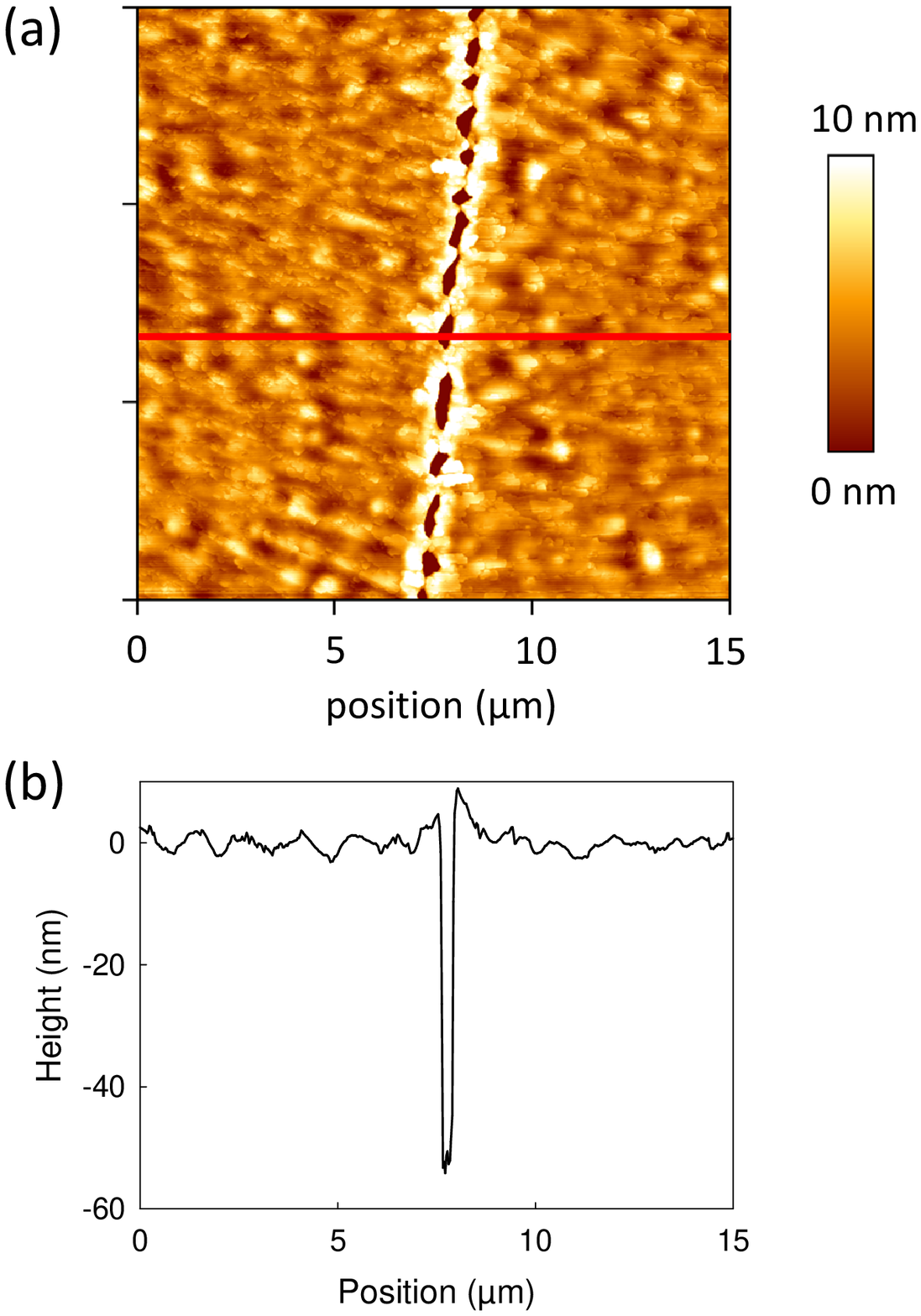}
\caption{(a) AFM image of a domain boundary between two platelet domains of an annealed SQIB film on glass. The cross section (b) along the red horizontal line in (a) provides the thickness of the film of approximately \SI{50}{nm}. }
\label{domainboundary}
\end{figure}

\clearpage

\begin{figure}[h]
\centering
\includegraphics[width=0.7\textwidth]{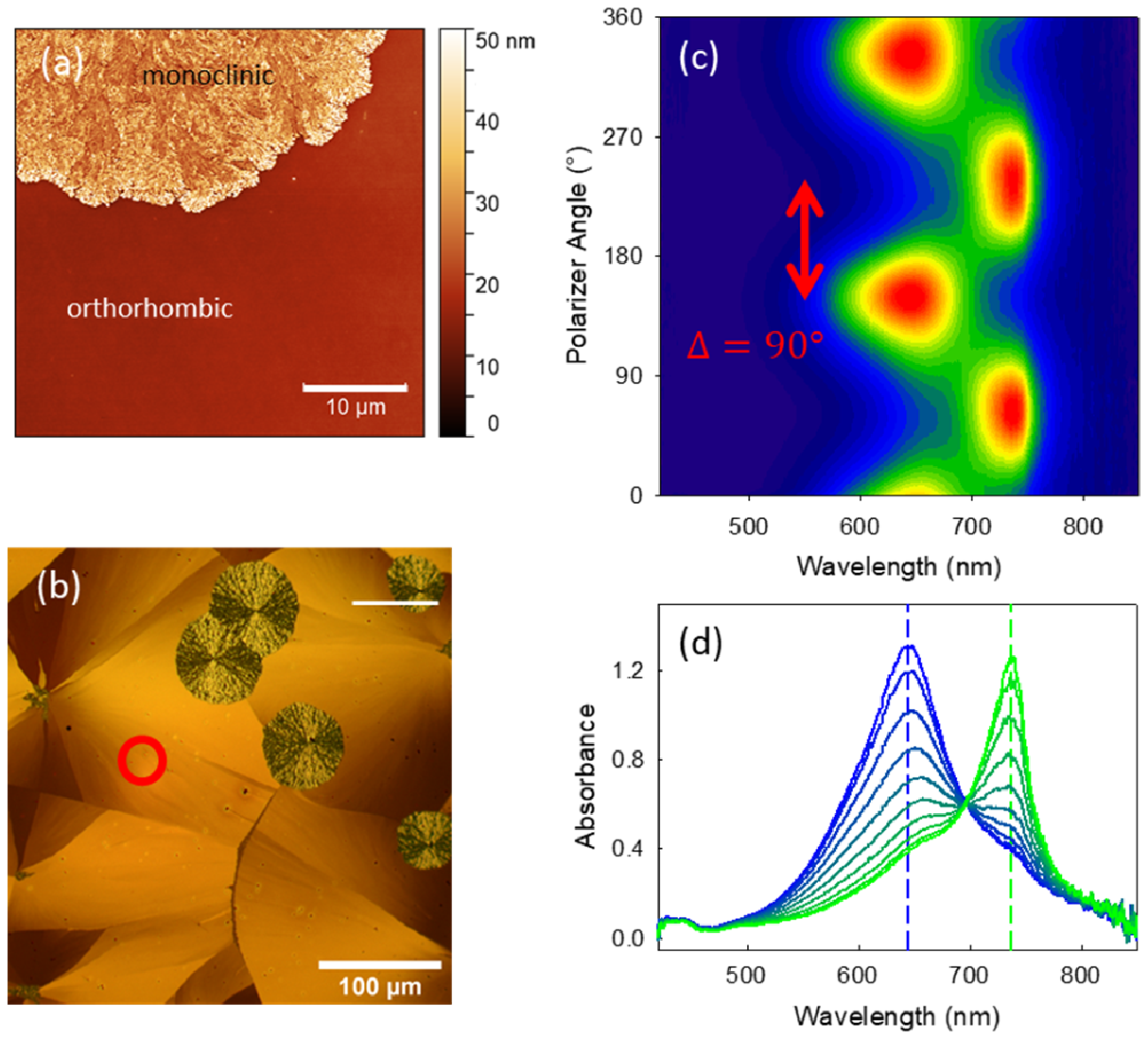}
\caption{The AFM image (a) elucidates the overall morphology of the two polymorphs of SQIB on glass. In (b), a polarized white light reflection microscopy image is shown. The red circle marks the area, where transmission spectra haven been taken: (c) Polarized absorbance spectra of SQIB platelets. In (d), single spectra are shown where the polarizer is rotated by \SI{10}{\degree} for every spectrum, covering \SI{90}{\degree} in total. The different polarization dependence of the upper (UDC) and lower Davydov (LDC) components is obvious. The dashed vertical lines mark the positions of the two maxima.  }
\label{polmic}
\end{figure}

Because of the domain size, polarized optical absorption spectroscopy is easiest done for the several \SI{10}{\micro m} large orthorhombic platelet domains, Figures~\ref{polmic}(a) and (b). The AFM image in Figure~\ref{polmic}(a) concurrently displaying both polymorph reveals the substantial roughness of the monoclinic phase due to the lacy texture. This small-sized texture impairs recording absorbance spectra on single monoclinic domains for the postannealed SQIB films on glass.

\begin{figure}[t]
\centering
\includegraphics[width=0.90\textwidth]{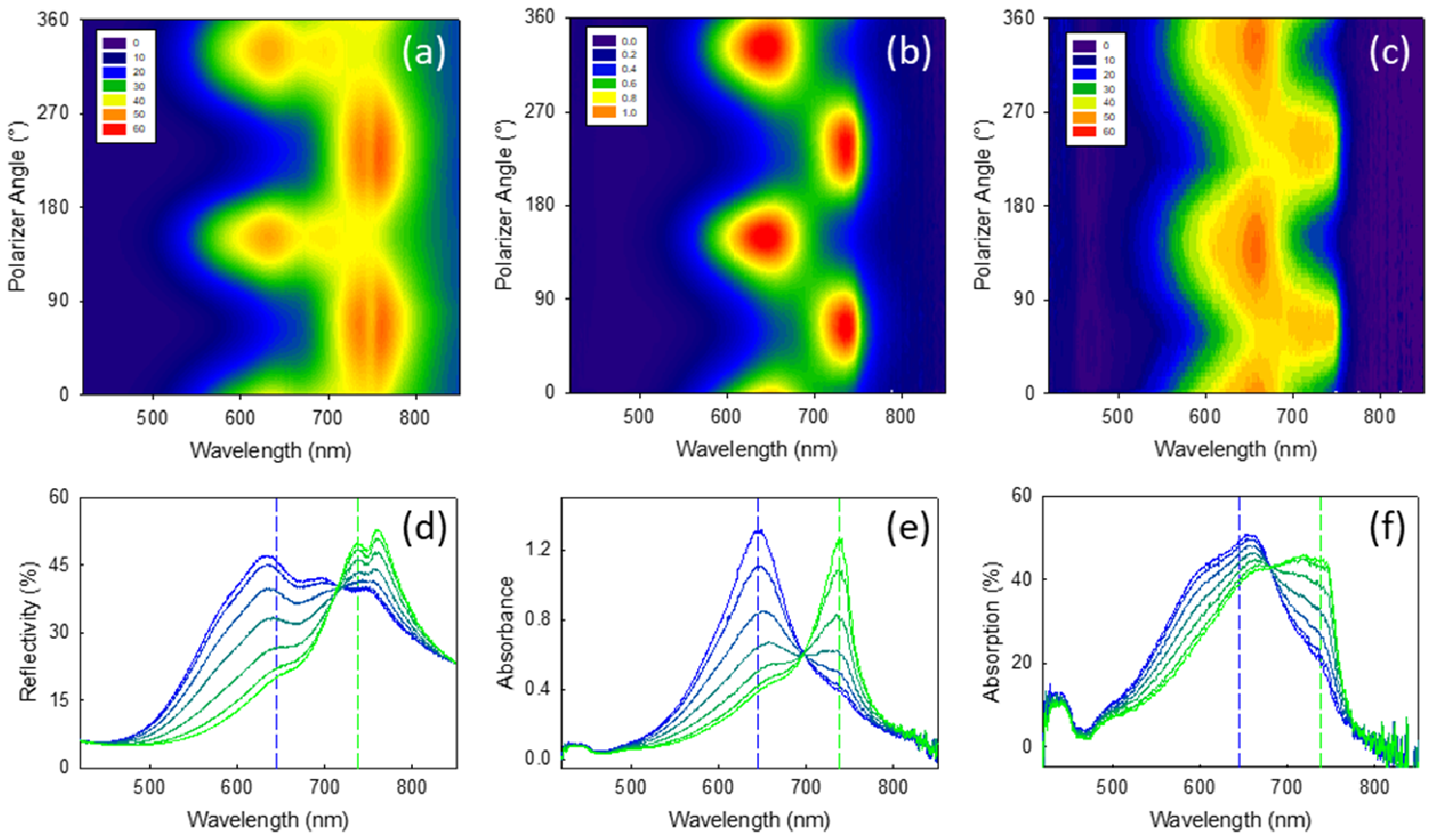}
\caption{The polarization-resolved specular reflectivity $R$, (a) and (d), absorbance $Abs=-\log(T)$, (b) and (e), calculated from the transmission $T$, and from this the calculated absorption $A=1-T-R$, (c) and (f), of a single SQIB platelet on glass. Illumination took place from the SQIB side. Vertical dashed lines mark the wavelengths of the two absorbance maxima. Maximal absorbance and reflectivity appear at the same polarisation angle.\cite{Balzer19}  }
\label{sqib_abs_refl}
\end{figure}

Even so the extended orthorhombic domains are not single crystalline, parts of them become completely dark when imaged between crossed polarizers.\cite{Balzer19,Funke21} In Figures~\ref{polmic}(c) and (d), absorbance spectra from part of a single orthorhombic platelet are shown. The position where the spectra have been taken is marked in the polarized reflection microscope image (b) by a red circle. Depending on the polarizer direction, either the UDC or LDC dominates the spectrum. As expected for the orthorhombic polymorph with the \hkl(110) plane parallel to the substrate, the angle $\Delta$ between the polarizer orientations for the two maxima is \SI{90}{\degree}. Note that the polarization angles for maximum absorbance also are the angles for maximum reflectivity, Figure~\ref{sqib_abs_refl}.

\clearpage

\begin{figure}[t]
\centering
\includegraphics[width=0.90\textwidth]{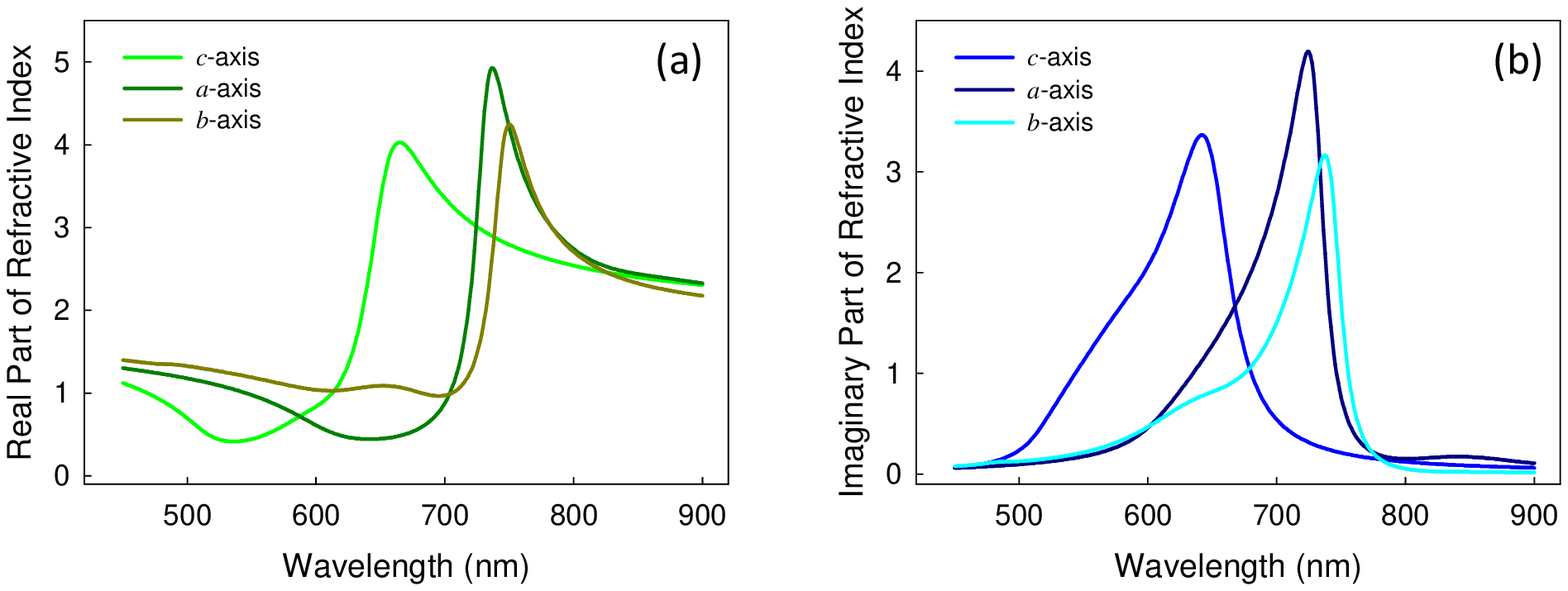}
\caption{Real (a) and imaginary (b) parts of the complex refractive index $N=n + i k$ along the three crystallographic axes of the orthorhombic SQIB polymorph.\cite{Funke21}}
\label{dielectrictensor}
\end{figure}

\clearpage

\begin{figure}[t]
\centering
\includegraphics[width=0.45\textwidth]{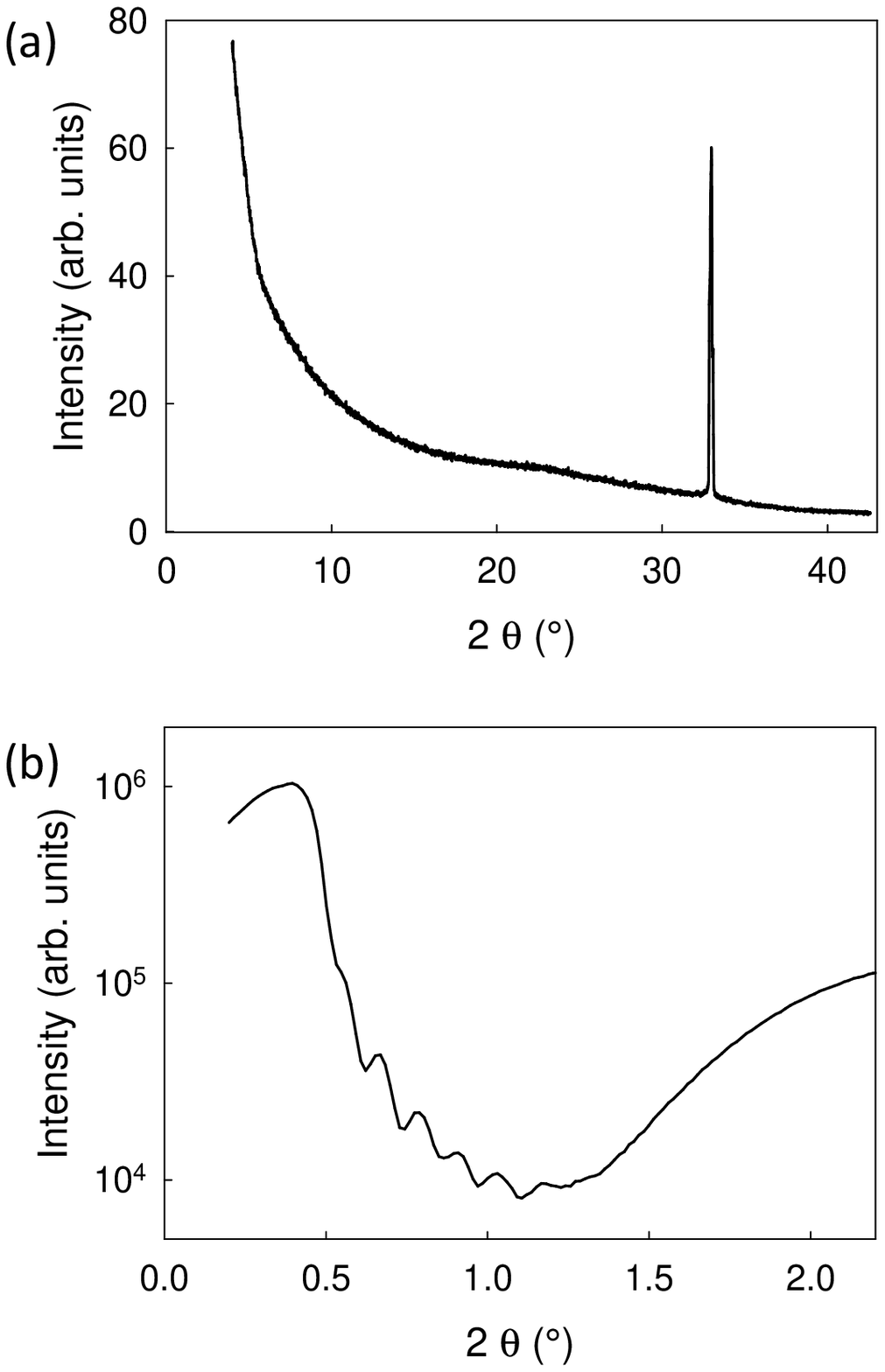}
\caption{(a) X-ray diffraction of an evaporated SQIB film on \ce{SiO2} (\ce{Si}-wafer covered with native oxide) with nominal thickness of \SI{30}{nm} (\ce{Cu} $K\alpha$ radiation $\lambda = \SI{1.541}{\angstrom}$). The diffractogram shows no peaks from the SQIB film --  the signal at $2\theta \approx \SI{33}{\degree}$ stems from the \ce{Si} wafer. (b) Well resolved Kiessig fringes prove the smoothness of the SQIB film. }
\label{kiessig}
\end{figure}

\clearpage

\begin{figure}[h]
\centering
\includegraphics[width=0.55\textwidth]{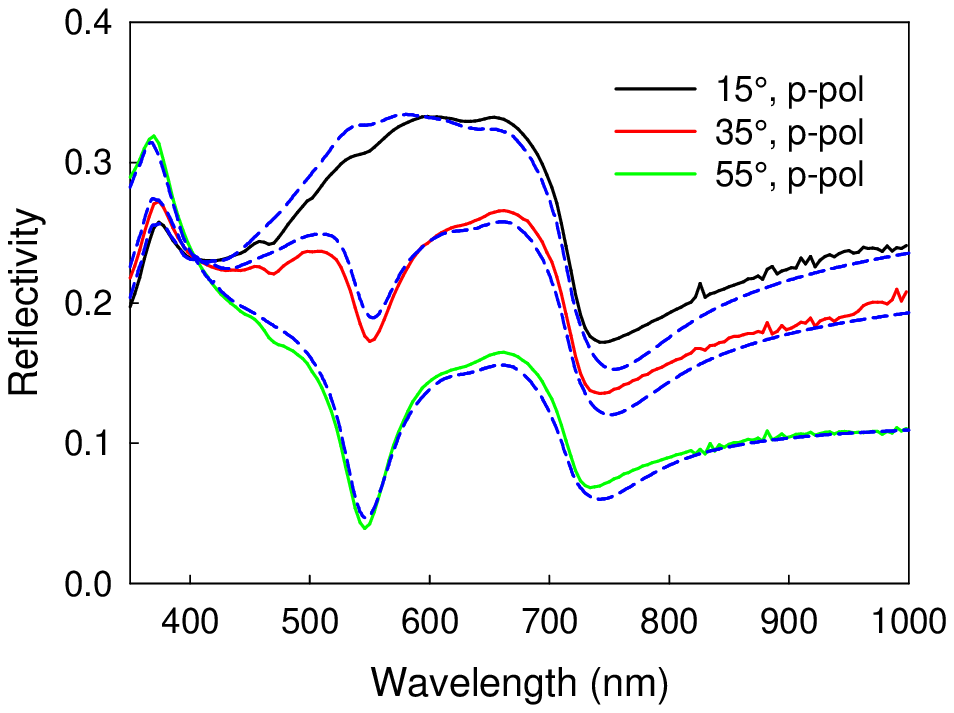}
\caption{Measured $p$-polarized reflection spectra (solid lines) of an evaporated SQIB film on \ce{SiO2} (\ce{Si}-wafer covered with \SI{2}{nm} native oxide) utilizing a J.A. Woollam rotating analyzer ellipsometer (VASE) with vertical sample stage. Dashed lines $p$-polarized reflection spectra based on the complex refractive index presented in Figure 3(b) in the main manuscript. The calculation was performed with OpenFilters setting the thickness of the SQIB layer to \SI{50}{nm}.\cite{Larouche08} }
\label{SQIB_refl_winkel}
\end{figure}

\clearpage

\begin{figure}[t]
\centering
\includegraphics[width=0.90\textwidth]{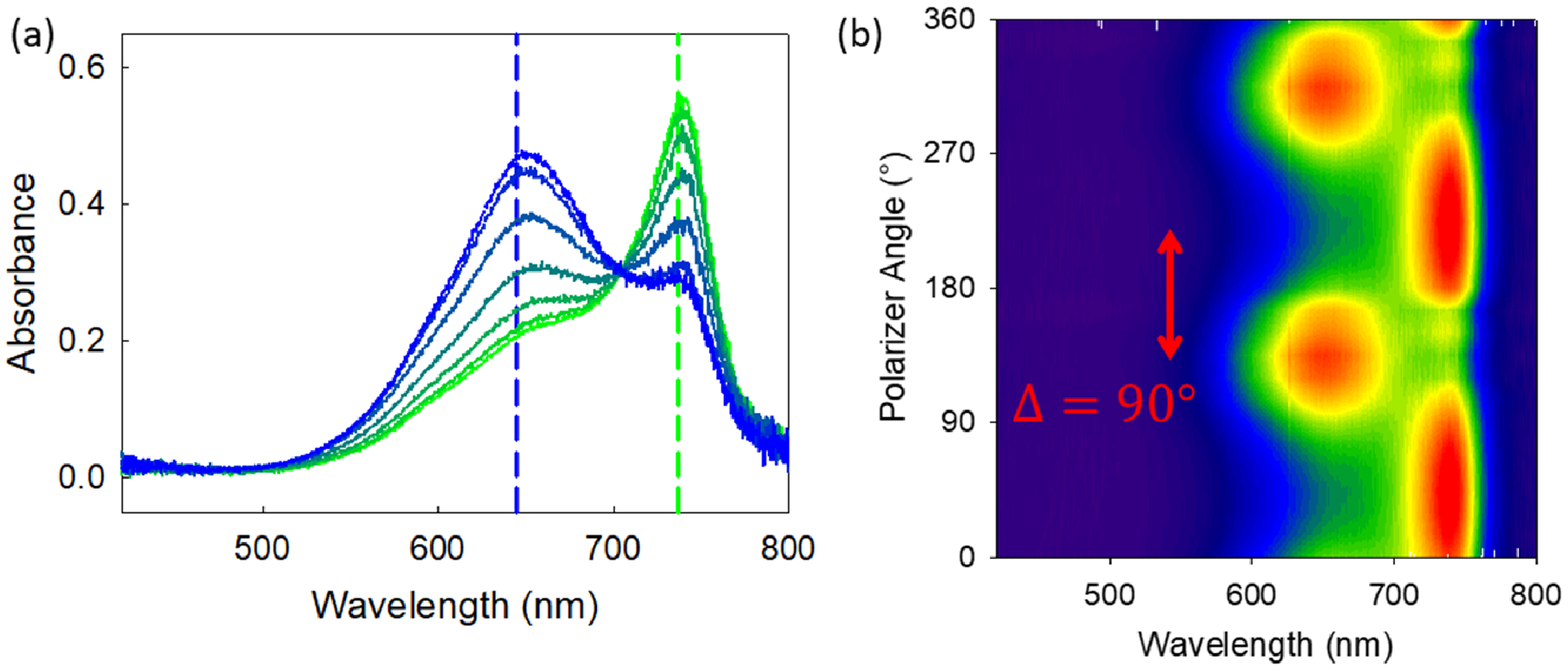}
\caption{Polarized absorbance spectra of SQIB, vacuum deposited on graphene/quartz. As reference, the bare substrate has been taken for each polarization direction. However, the birefringent quartz substrate leads to a broadening in the polarization angle dependence. The dashed, vertical lines mark the positions of the absorbance maxima for SQIB platelets on glass. }
\label{graphene_polarized}
\end{figure}

\clearpage

\begin{figure}[t]
\centering
\includegraphics[width=0.80\textwidth]{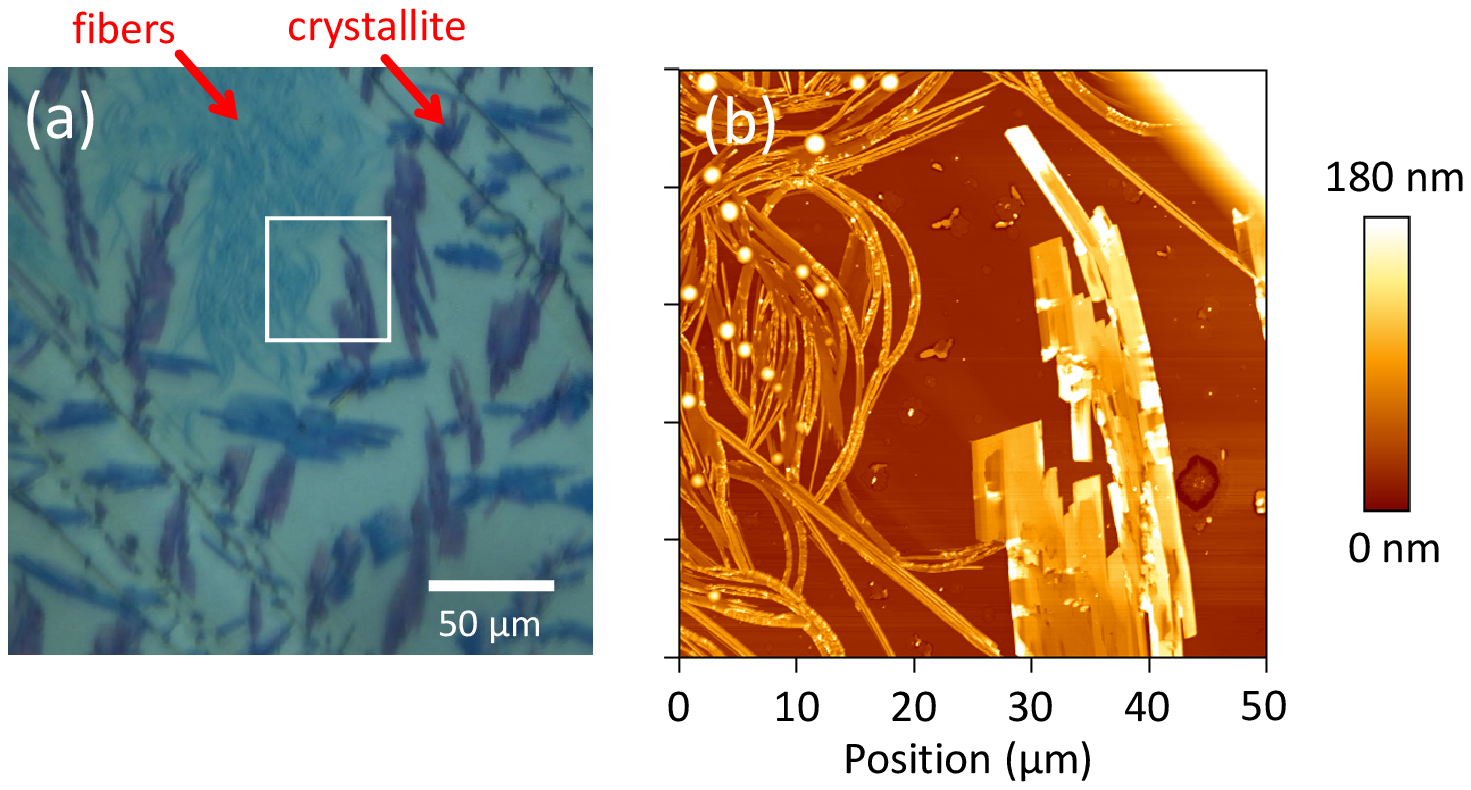}
\caption{Transmission optical microscope (a) and AFM image (b) of crystallites and fibers grown by vapor-deposition on \ce{KCl}. The white square in (a) marks the position of the AFM image (b). On top of some fibers, small droplets are often observed. Note, that different colors for the crystallites in (a) stem from a slight linear polarization of the illumination light.  }
\label{AFM_KCl}
\end{figure}


\begin{figure}[t]
\centering
\includegraphics[width=0.90\textwidth]{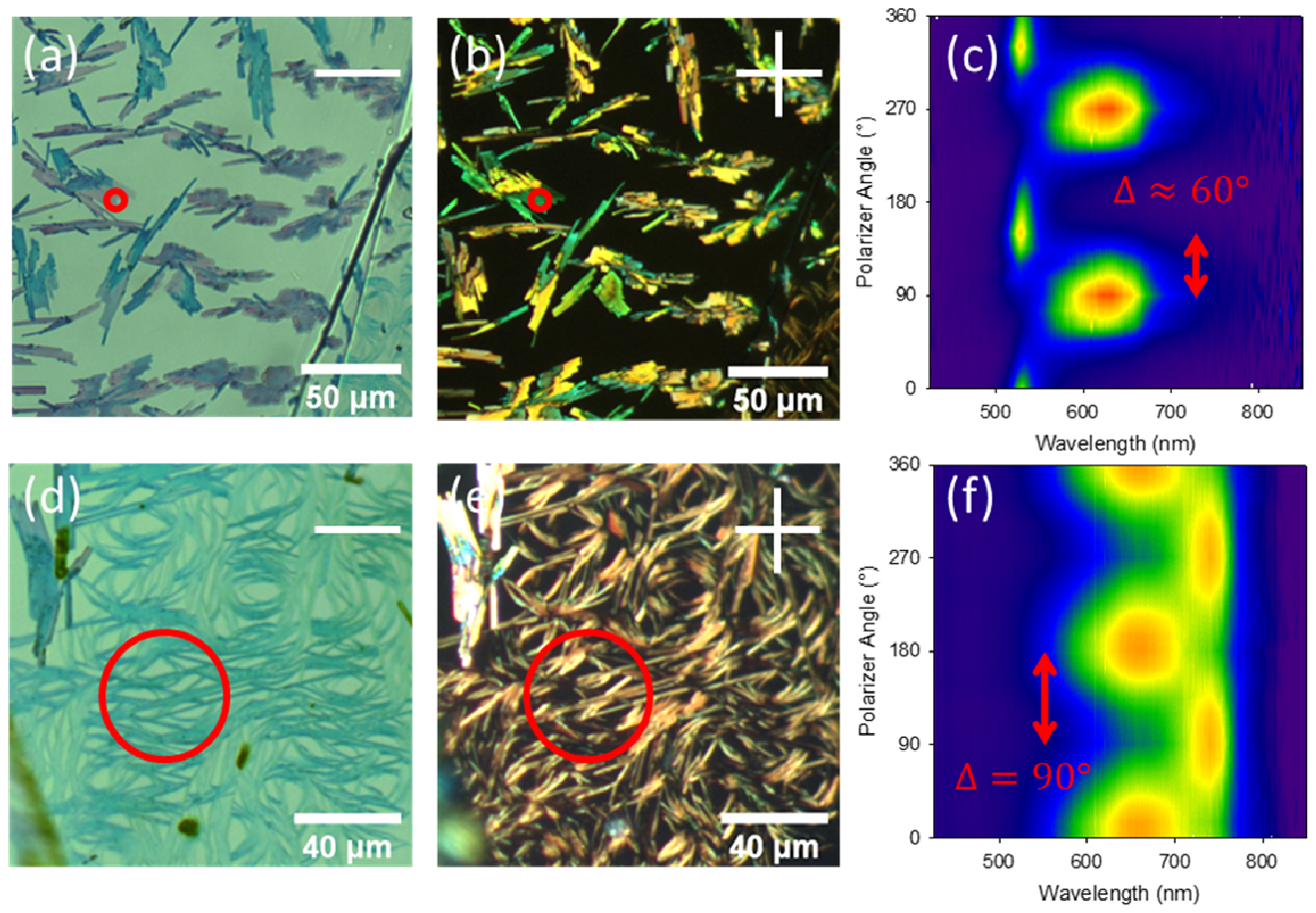}
\caption{Polarized transmission (single horizontal polarizer) and reflection (crossed polarizers, horizontal and vertical) microscope images of monoclinic crystallites (a,b) and orthorhombic fibers (d,e) of SQIB formed on \ce{KCl}\hkl(001). The corresponding polarized absorbance spectra are shown in (c) and (f), taken at the positions marked by the red circles. The angle $\Delta$ provides the angular difference for the absorbance maxima. }
\label{sqib_KCl_polari}
\end{figure}

\clearpage

\begin{figure}[t]
\centering
\includegraphics[width=0.85\textwidth]{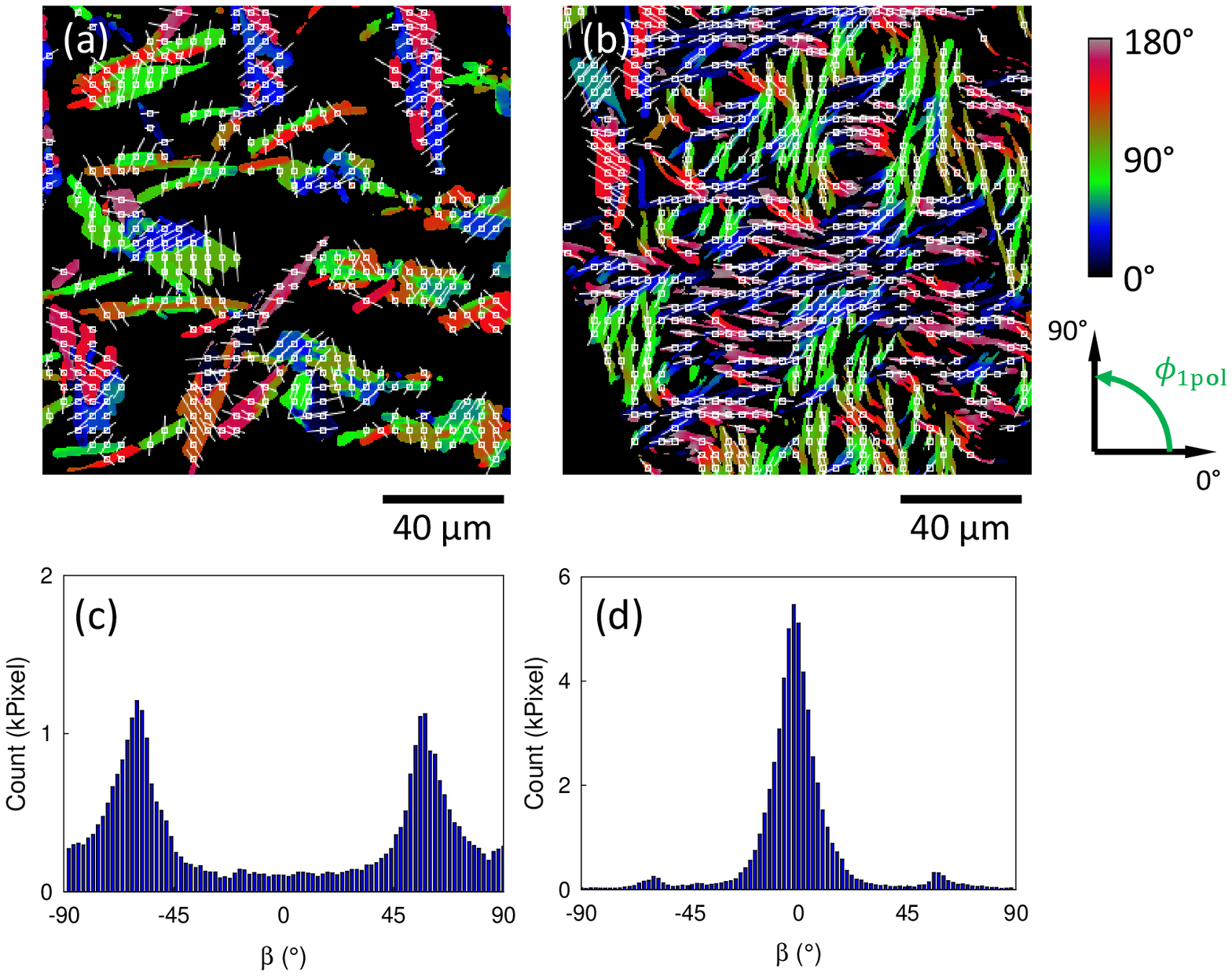}
\caption{(a) Polarization analysis for the monoclinic SQIB crystallites formed on \ce{KCl}, Figure~\ref{sqib_KCl_polari}. For this, the wavelength $\lambda=\SI{650}{nm}$ has been chosen, corresponding to the UDC. The white lines together with the color code mark the polarization angle $\phi_{\mathrm{1pol}}$ of maximum reflectivity.  (b) The same for the LDC at \SI{650}{nm} for the orthorhombic phase. In (c) and (d), the angles of maximum reflectivity with respect to the long crystallite direction, $\beta = \phi_{\mathrm{1pol}}-\theta_{\mathrm{orient}}$, are plotted. The angle $\beta$ is zero if the maximum reflectivity coincides with light polarized along the long crystallite or fiber direction. For the monoclinic crystallites, the histogram in (c) shows peaks at $|\beta| = \SI{58\pm 5}{\degree}$. For all of the orthorhombic fibers, $\beta = \SI{0\pm 5}{\degree}$.   }
\label{KCl_Richtungen}
\end{figure}

\clearpage


%